\documentclass[11pt,a4paper, english, iop, numberedappendix]{emulateapj}

\usepackage[T1]{fontenc}
\usepackage[latin1]{inputenc}
\usepackage{graphicx}
\usepackage{amssymb}

\usepackage{times}
\usepackage{amsmath}

\newcommand{\lya}{Ly$\alpha$}

\slugcomment{}

\shorttitle{HI near galaxies at $z\approx2.4$}
\shortauthors{Rakic et al.}

\begin{document}


\title{Neutral hydrogen optical depth near star-forming galaxies at $z\approx2.4$ in the Keck Baryonic Structure Survey\altaffilmark{*}} 



\author{Olivera Rakic, Joop Schaye,}
\affil{Leiden Observatory, Leiden University,  P.O. Box 9513, 2300 RA Leiden, The Netherlands}
\author{Charles~C.~Steidel, and Gwen~C.~Rudie}
\affil{California Institute of Technology, MS 249-17, Pasadena, CA 91125, USA}

\altaffiltext{*}{Based on data obtained at the W.M. Keck Observatory, which is operated as a scientific partnership among the California Institute of Technology, the University of California, and NASA, and was made possible by the generous financial support of the W.M. Keck Foundation.}







\begin{abstract}
We study the interface between galaxies and the intergalactic medium
by measuring the absorption by neutral hydrogen in the vicinity of star-forming galaxies at $z\approx 2.4$. Our sample consists of 679 rest-frame-UV selected galaxies with spectroscopic redshifts that have impact parameters $< 2$ (proper) Mpc to the line of sight of one of 15 bright, background QSOs and that fall within the redshift range of its \lya\ forest. We present the first 2-D maps of the absorption around galaxies, plotting the median \lya\ pixel optical depth as a function of transverse and line of sight separation from galaxies. The \lya\ optical depths are measured using an automatic algorithm that takes advantage of all available Lyman series lines. The median optical depth, and hence the median density of atomic hydrogen, drops by more than an order of magnitude around 100 kpc, which is similar to the virial radius of the halos thought to host the galaxies. The median remains enhanced, at the $>3\sigma$ level, out to at least 2.8 Mpc (i.e.\ $> 9$ comoving Mpc), but the scatter at a given distance is large compared with the median excess optical depth, suggesting that the gas is clumpy.
Within 100 (200) kpc, and over $\pm165\rm\, km\, s^{-1}$, the covering fraction of gas with \lya\ optical depth greater than unity is $100^{+0}_{-32}\%$ ($66\pm 16\%$). Absorbers with $\tau_{\rm Ly\alpha}> 0.1$ are typically closer to galaxies than random. The mean galaxy overdensity around absorbers increases with the optical depth and also as the length scale over which the galaxy overdensity is evaluated is decreased. Absorbers with $\tau_{\rm Ly\alpha}\sim 1$ reside in regions where the galaxy number density is close to the cosmic mean on scales $\ge 0.25$ Mpc. We clearly detect two types of redshift space anisotropies. On scales $< 200 ~\rm km\,s^{-1}$, or $< 1$ Mpc, the absorption is stronger along the line of sight than in the transverse direction. This ``finger of God'' effect may be due to redshift errors, but is probably dominated by gas motions within or very close to the halos. On the other hand, on scales of 1.4 - 2.0 Mpc the absorption is compressed along the line of sight (with $>3\sigma$ significance), an effect that we attribute to large-scale infall (i.e.\ the Kaiser effect). 
\end{abstract}


\section{Introduction}
Gas accretion and galactic winds are two of the most important and poorly understood ingredients of models for the formation and evolution of galaxies. One way to constrain how galaxies get their gas, and to learn about the extent of galactic feedback, is to study the intergalactic medium (IGM) in the galaxies' vicinity. The interface between galaxies and the IGM can be studied either in emission \citep[e.g.][]{Bland1988,Lehnert1999,Ryan-Weber2006,Borthakur2010,Steidel2011} or in absorption against the continuum of background objects such as QSOs \citep[e.g.,][]{
Lanzetta1990, Bergeron1991, Steidel1992, Steidel1994, Lanzetta1995, Steidel1997, Chen1998, Chen2001, Bowen2002, Penton2002, Frank2003, Adelberger2003, Adelberger2005, Pieri2006, Simcoe2006, Steidel2010, Crighton2011, Prochaska2011, Kacprzak2011, Bouche2011} or galaxies \citep[e.g.][]{Adelberger2005, Rubin2009, Steidel2010, Bordoloi2011}. 

Emission from intergalactic gas is very faint and observations are currently mostly limited to low redshifts. At $z\approx 2.7$ Ly$\alpha$ emission was recently seen out to $\sim80$ kpc from star-forming galaxies in a stacking analysis by \citet{Steidel2011}, and the origin of this light seems to be radiation of the central object scattered by galactic halo gas. However, these observations are limited to the immediate  vicinity of galaxies, i.e.\ $\approx 100\rm \, kpc$, and are currently feasible only for the \ion{H}{1} Ly$\alpha$ transition. On the other hand, studying this gas in absorption is viable at all redshifts as long as there are sufficiently bright background objects. Most importantly, absorption studies are sensitive to gas with several orders of magnitude lower density than emission studies.  

The background sources used for absorption line probes of the IGM have traditionally been QSOs, but one may also use background galaxies or gamma ray bursts. While the surface density of background galaxies is much higher than that of QSOs, the quality of the individual spectra is much lower, since at the redshifts discussed in this paper the typical background galaxy is $>10^3$ times fainter than the brightest QSOs. Studies using galaxies as background sources are therefore  confined to analyzing strong absorption lines and generally require stacking many lines of sight.
On the other hand, QSOs sufficiently bright for high-resolution, high S/N spectroscopy using 8m-class telescopes are exceedingly rare, but the information obtained from a single line of sight is of exceptional quality, well beyond what could be obtained with even the very brightest galaxy at comparable redshift. We note also that galaxies and QSOs do not provide identical information, as the former are much more extended than the latter. This is particularly relevant for metal lines, which often arise in absorbers with sizes that are comparable or smaller than the half-light radii of galaxies \citep[e.g.][]{Schaye2007}. 

In this paper we focus on studying the IGM near star-forming galaxies at $2.0 < z < 2.8$ using absorption spectra of background QSOs. At present we focus on \ion{H}{1} \lya\ in the vicinity of galaxies, while a future paper will study the relation between metals and galaxies. Star-forming galaxies in this redshift range can be detected very efficiently based on their rest-frame UV colors \citep{Steidel2003,Steidel2004} and the same redshift range is ideal for studying many astrophysically interesting lines that lie in the rest-frame UV part of the spectrum (e.g.\, Ly$\alpha$ at 1215.67\AA, CIV $\lambda \lambda$ 1548.19,1550.77 \AA, OVI  $\lambda \lambda$1031.93,1037.616 \AA, where the CIV and OVI lines are doublets). At these redshifts the Ly$\alpha$ forest is not as saturated as at $z\gtrsim 4$, and at the same time the absorption systems are not as rare as they are at low redshifts. In addition, this redshift range is exceptionally well-suited for studying the galaxy-IGM interface given that this is when the universal star formation rate was at its peak \citep[e.g.][]{Reddy2008}, and hence we expect the interaction between galaxies and their surroundings to be most vigorous. 

The largest  QSO-galaxy surveys at high redshift are those from \citet{Adelberger2003,Adelberger2005} and \citet{Crighton2011}, and here we will briefly describe the former given that it is the most comparable in terms of the galaxy sampling density and data quality to the study presented in this paper.   \citet{Adelberger2003} studied the IGM close to 431 Lyman Break Galaxies (LBGs) at $z\approx3-4$ in 8 QSO fields. They found enhanced Ly$\alpha$ absorption within $\approx1-6\, h^{-1}$ comoving Mpc ($\approx0.3-2.1$ proper Mpc) from star-forming galaxies. On the other hand, they found that the region within $\lesssim0.5\, h^{-1}$ comoving Mpc ($\approx 0.2$ proper Mpc) contains less neutral hydrogen than the global average. This last result was, however, based on only 3 galaxies. \citet{Adelberger2005} studied the IGM at $1.8\lesssim z\lesssim3.3$ with an enlarged sample: 23 QSOs in 12 fields containing 1044 galaxies. They confirmed the earlier result of enhanced absorption within $\lesssim7h^{-1}$ comoving Mpc ($\approx2.4$ proper Mpc) and found that even though most galaxies show enhanced absorption within $\lesssim0.5\, h^{-1}$ comoving Mpc, a third of their galaxies did not have significant associated Ly$\alpha$ absorption. 

In comparison with \citet{Adelberger2003,Adelberger2005} we use
uniformly excellent QSO spectra taken with the HIRES echelle spectrograph, covering from $\simeq$3100\,\AA\, to
at least the QSO's CIV emission line. \citet{Adelberger2003} also used HIRES spectra, but at $z>3$ the surface density of galaxies with $R_{\rm s}<25.5$ is $\sim4-5$ times smaller than at $z=2.4$, and the QSO spectra did not cover higher order Lyman lines, which are important for the recovery of optical depths in saturated Ly$\alpha$ lines. \citet{Adelberger2005} used HIRES spectra as well as lower-resolution spectra of fainter QSOs obtained using LRIS and ESI.

This paper is organized as follows. We describe our data sample in Section~\ref{Data}. In Section~\ref{POD} we discuss the so-called pixel optical depth method that we use to analyze the QSO spectra. The distribution of Ly$\alpha$ absorption as a function of transverse and line of sight (LOS) separation from galaxies is presented in Section~\ref{Results}, while the distribution of galaxies around absorbers is described in Section~\ref{IGMCentricView}. Finally, we conclude in Section~\ref{Summary}.  

Throughout this work we use $\Omega_{\rm m}=0.258$, $\Omega_{\Lambda}=0.742$, $\Omega_{\rm b}=0.0418$, and $h=0.719$ \citep{Komatsu2009}. When referring to distances in proper (comoving) units we denote them as pMpc (cMpc).

\section{Data}\label{Data}

\begin{deluxetable*}{@{}lllllrrrr@{$\times$}rc}
\tabletypesize{\scriptsize}
\tablecaption{QSO fields. 
\label{tbl-1}}
\tablewidth{0pt}
\tablehead{
\colhead{QSO\tablenotemark{a}}  & \colhead{$z$\tablenotemark{b}} & \colhead{$G_{AB}$\tablenotemark{c}}  & \colhead{$z_{\rm min}$\tablenotemark{d}} & \colhead{$z_{\rm max}$}\tablenotemark{e} & \colhead{$\lambda_{\rm min}$\tablenotemark{f}}  [\AA] & \colhead{$\lambda_{\rm max}$\tablenotemark{g}}  [\AA]& \colhead{S/N$_{\rm HI}$\tablenotemark{h}}  & \colhead{$\Delta \Omega$\tablenotemark{i}}& &\colhead{N$_{\rm gal}$\tablenotemark{j}} 
}
\startdata
Q0100+13  	&   2.7210  &   	16.6  &  2.196	&  2.654	&	3125		&	8595 	&	67 	&	$5.6'$& $7.6'$ 	& 35\\ 
Q0105+1619	&   2.6518	 &	16.9	 &  2.102	&  2.603	&	3225		&	6041		&	112	&	$5.4'$& $7.4' $ 	& 59\\
Q0142-09    	&   2.7434	 &   	16.9  &  2.210	&  2.671	&	3100		& 	6156		&	65 	& 	$5.4'$& $7.4'$ 	& 45\\
Q0207-003	&   2.8726	 &	16.7   & 2.264	&  2.848	&	3100		&	9220		&	71	&	$5.4'$& $7.0'$ 	& 38\\
Q0449-1645 	&   2.6840	 & 	17.0   & 2.094	&  2.651	&	3150		& 	5966 	&	67 	&	$5.0'$& $6.5'$	& 51\\
Q0821+3107 	&   2.6156  &  	17.3   & 2.150	&  2.584	&	3235		& 	5967 	& 	42	&	$5.4'$& $7.4' $ 	& 32\\
Q1009+29 	&   2.6520  & 	16.0   & 2.095	&  2.593	&	3160		&	7032		& 	87 	&	$5.2'$&  $7.2' $ & 29\\
Q1217+499 	&   2.7040	 &   	17.1   & 2.122	&  2.680	&	3075		& 	7000 	& 	64 	&	$5.1'$& $6.9' $ 	& 37\\
Q1442+2931	&   2.6603	 &	17.0   & 2.078	&  2.638	&	3155		&	6150		&	83 	&	$5.4'$& $7.5' $ 	& 44\\
Q1549+1933	&   2.8443 &	16.3   & 2.238	&  2.814	&	3160		&	9762		&	164  &	$5.2'$& $7.1' $ 	& 44\\
HS1603+3820	&   2.5510	 &	15.9   & 1.974	&  2.454	&	3185		&	9762		&	87 	&	$5.4'$& $7.2' $ 	& 41\\
Q1623-KP77  	&   2.5353  &  	17.4   & 1.983	&  2.505	&	3125		& 	6075 	& 	43     &	$16.1'$& $11.6' $ & 54\\
Q1700+64    	&   2.7513 &   	16.1  & 2.168	&  2.709	&	3145		& 	9981		& 	92 	&	$11.5'$& $11.0' $ & 64\\
Q2206-199   	&   2.5730 & 	17.5  & 2.009	&  2.541	&	3047		& 	10088	& 	73 	&	$5.4' $ & $7.5' $ & 46\\
Q2343+12 	&   2.5730	 &   	17.0  & 2.012	&  2.546	&	3160		& 	10087 	&	57 	&	$22.5'$& $8.5' $ & 60\\
\enddata
\tablenotetext{a}{QSO name}
\tablenotetext{b}{QSO redshift}
\tablenotetext{c}{QSO AB magnitude (Sloan g' band)}
\tablenotetext{d}{Minimum galaxy redshift}
\tablenotetext{e}{Maximum galaxy redshift}
\tablenotetext{f}{Minimum wavelength covered}
\tablenotetext{g}{Maximum wavelength covered}
\tablenotetext{h}{Median S/N per pixel in the analyzed Ly$\alpha$ region}
\tablenotetext{i}{Spectroscopically observed area}
\tablenotetext{j}{Number of galaxies with spectroscopic redshifts}
\end{deluxetable*}

\subsection{The Keck Baryonic Structure Survey}\label{galaxies}

The Keck Baryonic Structure Survey (KBSS; Steidel et al 2012) is a new survey
which combines high precision studies of the IGM with targeted galaxy redshift
surveys of the surrounding volumes, expressly designed to establish
the galaxy/IGM connection in the redshift range $1.8 \lesssim z \lesssim 3$.
The KBSS fields are centered on 15 background QSOs (see Table~\ref{tbl-1}) that are among
the most luminous known (${\rm L_{bol} \gtrsim 10^{14}}$ L$_{\odot}$) in the redshift range
$2.5 \lesssim z \lesssim 2.85$.  The QSO redshifts were chosen to maximize the information
content and redshift path sampled by their absorption line spectra. Toward that end,
the QSO spectra used in the KBSS are of unprecedented quality, combining archival
high resolution spectra from Keck/HIRES (and, in 3 cases, archival VLT/UVES spectra
as well) with new Keck/HIRES observations to produce the final co-added spectra
summarized in Table~\ref{tbl-1}. 

The mean flux in this sample of QSO spectra is 0.806. The formula for the effective Ly$\alpha$ optical depth from \citet{Schaye2003} suggests that the mean flux at $z=2.36$ is 0.802($\pm0.008$), which is
consistent with our data, so we conclude that the IGM probed by QSO
sightlines in our sample is representative for the considered redshift.

Within each KBSS
field, UV color selection techniques (\citealt{Steidel2003,Adelberger2004}) were used
to tune the galaxy redshift selection
function so as to cover the same optimal range of redshifts probed by the QSO spectra.
The spectroscopic follow-up using Keck/LRIS was carried out over relatively small
solid angles surrounding the QSO sightlines, but typically 8 to 10 slit masks with
nearly identical footprint were obtained in each field, leading to a high level of
spectroscopic completeness and a very dense sampling of the survey volumes surrounding
the QSO sightlines.
The full KBSS galaxy sample contains 2188 spectroscopically identified galaxies in the redshift
range $1.5 \le z \le 3.5$, with $\langle z \rangle = 2.36\pm0.42$.
The spectroscopic sample was limited to those with apparent magnitude
${\cal R} \le 25.5$, where galaxies in the range $23 \le {\cal R} \le 24.5$ and
those within 1 arcminute of a QSO sightline were given highest priority in designing
slit masks.

The effective survey area is $720\rm\, arcmin^2$ (i.e.\, 0.2
square degrees), and the average surface density of galaxies with
spectroscopic redshifts is $3.1\rm\, arcmin^{-2}$ (within 2.5 pMpc from
QSO sightlines).

In this paper, we
focus on a subset of 679 galaxies that satisfy the following criteria for redshift
and projected distance from the relevant QSO sightline:
\begin{itemize}
\item[1)] their redshifts are within the Ly$\alpha$ forest range, defined as:
\begin{equation}
(1+z_{\rm QSO})\frac{\lambda_{\rm Ly\beta}}{\lambda_{\rm Ly\alpha}}-1< z < z_{\rm QSO}-(1+z_{\rm QSO})\frac{3000~ \rm km\, s^{-1}}{c}, \label{eq:zrange}
\end{equation}

where  $z_{\rm QSO}$ is the QSO redshift, and $\lambda_{\rm Ly\alpha}$ and $\lambda_{\rm Ly\beta}$ are the rest-frame  wavelengths of the hydrogen Ly$\alpha$ (1215.67\AA) and  Ly$\beta$ (1025.72\AA) lines,  respectively (for a discussion of the limits in this expression, see \S\ref{QSOspectra});  
\item[2)] they are within 2 pMpc ($\sim 4$\arcmin\ or  $\sim 5 h^{-1}$ cMpc at $z = 2.4$), the transverse
distance to which there is significant coverage in all 15 KBSS fields.
\end{itemize}
Figure 1 shows the number of galaxies  as a function of
their (proper) distance from the QSO sightlines.

We note that 3 of the 15 KBSS fields (1623+268, 1700+64, 2343+12)
were also included in the study by Adelberger et al (2005).
However, since 2005 the data have been increased in both quantity and quality (for both
the QSO spectra and the galaxy surveys) in all 3 of these fields. Further details on the
KBSS survey and its data products will be described by Steidel et al (2012).

The smallest impact parameter in the present sample
is 55 pkpc, with  29, 106, and 267 galaxies
having impact parameters smaller than 200, 500, and 1000 pkpc,
respectively.

\begin{figure}
\epsscale{1.2}
\plotone{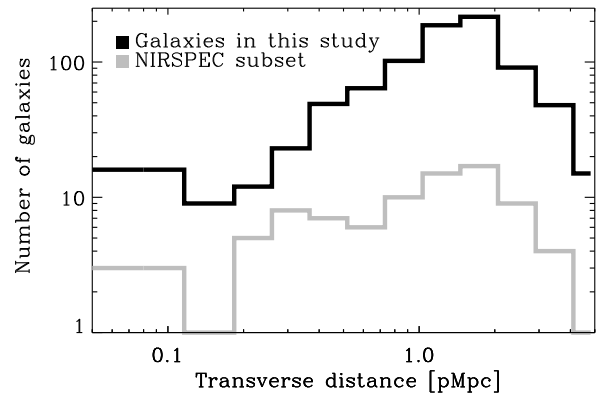}
\caption{Number of galaxies as a function of proper distance from the line of sight to the background QSO. We do not use galaxies with impact parameters larger than 2 pMpc since such galaxies were only targeted for a fraction of our fields. The distance bins are chosen to match the bins that are used for most of the figures in this paper (bins are separated by 0.15 dex). }\label{distance}
\end{figure}

\subsubsection{Redshifts}\label{Redshifts}
The redshifts of the vast majority (608 out of 679) of the galaxies in our sample are derived from interstellar absorption lines 
and/or the Ly$\alpha$ emission line. These lines were measured from low-resolution
(FWHM $\approx370\,\rm km\, s^{-1}$) multislit spectra taken between
2000 and 2010 with LRIS-B on the Keck I and II telescopes. However, ideally one would want to measure the galaxy redshifts from the nebular emission lines  ([\ion{O}{2}] $\lambda 3727$, H$\alpha$, H$\beta$, [\ion{O}{3}] $\lambda\lambda 4959,5007$) since those originate in stellar \ion{H}{2} regions and are more likely to correspond to the galaxies' systemic redshifts. \citet{Erb2006b} have used NIRSPEC, a near-IR instrument on  Keck II, to obtain higher resolution  (FWHM $\approx 240~\rm km\, s^{-1}$) spectra than achieved by LRIS-B and have used these to measure nebular redshifts for 110 galaxies. Our sample contains 71 of those galaxies and we take their systemic redshifts to be equal to the nebular redshifts, i.e.\ $z_{\rm gal}=z_{\rm neb}$. Marginally resolved lines have $\sigma\approx160~\rm km\, s^{-1}$ for
LRIS-B and $\sigma\approx100~\rm km\, s^{-1}$ for NIRSPEC.
\begin{figure}
\epsscale{1.2}
\plotone{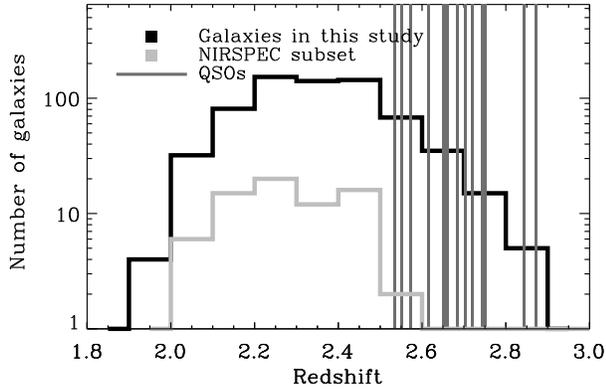}
\caption{Redshift distribution of QSOs (vertical lines) and galaxies that are used in our study, i.e.\ of galaxies that are in the Lyman-$\alpha$ forest redshift ranges of QSOs in their fields and that have impact parameters $<2$ pMpc. The black histogram shows the redshifts of all galaxies (679 objects in total), and the grey histogram just those with redshifts measured from near-IR spectra (71 galaxies).}\label{redshift}
\end{figure}

For the galaxies without near-IR observations we estimate the systemic redshifts using the empirical relations of \citet{Rakic2011}. Using the same data as analyzed here, they calibrated galaxy redshifts measured from rest-frame UV lines (Lyman-$\alpha$ emission, $z_{\rm Ly\alpha}$, and interstellar absorption, $z_{\rm ISM}$) by utilizing the fact that the mean  Ly$\alpha$ absorption profiles around the galaxies, as seen in spectra of background QSOs, must be symmetric with respect to the true galaxy redshifts if the galaxies are oriented  randomly with respect to the lines of sight to the background objects. The following values represent their best fits to the data:
\begin{enumerate}
\item[a)] for galaxies without available interstellar absorption lines (70 objects):
\begin{equation}
z_{\rm gal}=z_{\rm Ly\alpha}-295_{-35}^{+35}\rm \, km\, s^{-1}
\end{equation}
\item[b)] for galaxies without detected Ly$\alpha$ emission lines (346 objects):
\begin{equation}
z_{\rm gal}=z_{\rm ISM} + 145_{-35}^{+70}\rm\, km\, s^{-1}
\end{equation}
\item[c)] for galaxies for which both interstellar absorption and  Ly$\alpha$ emission are detected (263 objects) we use the arithmetic mean of the  above expressions. 
\end{enumerate}

These offsets yield redshift estimates free of velocity systematics \citep{Rakic2011, Rudie2012}. Repeated observations of the same galaxies suggest that the typical measurement uncertainties are   $ \approx 60~\rm km\, s^{-1}$ for  $z_{\rm neb}$. Comparison of redshifts inferred from rest-frame UV lines with those from nebular lines for the subset of 89 galaxies that have been observed in the near IR, shows that the rest-frame UV inferred redshifts have a  scatter of $\approx 130\, \rm km\, s^{-1}$. In addition, \citet{Rudie2012} concluded that redshift errors in the KBSS sample are $\lesssim 160\, \rm km\, s^{-1}$ based on the velocity structure of the strongest \ion{H}{1} absorbers around galaxy positions (first panel of their Figure 18), and \citet{Trainor2012} reached the same conclusion based on the small observed velocity dispersion of galaxies with respect to QSOs.

Figure~\ref{redshift} shows a histogram of the galaxy redshifts, both for the full sample and for the subsample with  redshifts measured from near-IR lines. The median redshifts of the full and near-IR samples are 2.36 and 2.29, respectively. The mean and standard deviations of the redshift distributions for the two samples are $2.36 \pm 0.17$ and $2.28 \pm 0.13$, respectively.

\begin{figure*}
\epsscale{1.}
\plotone{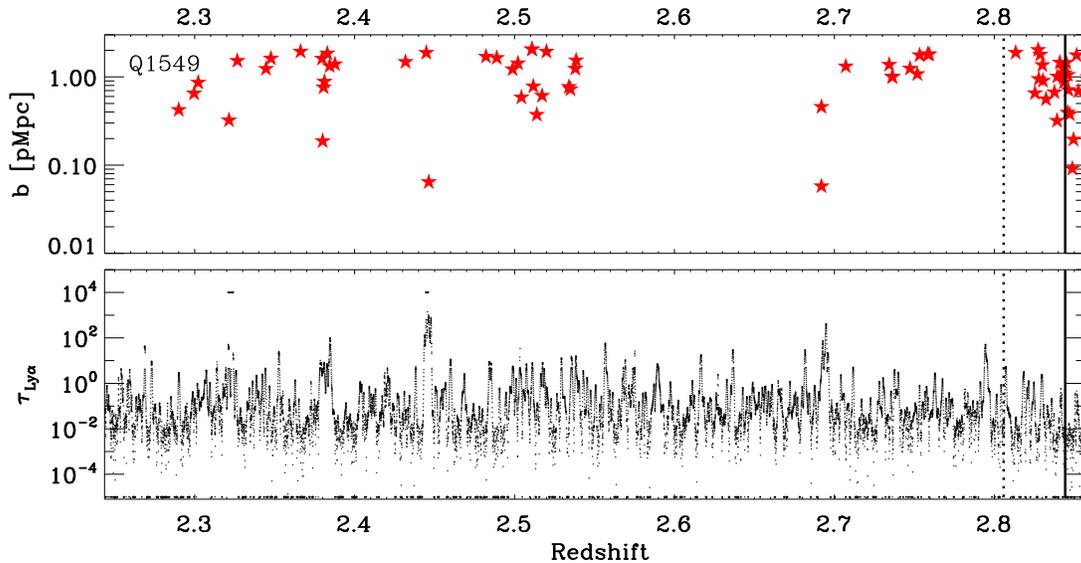}
\caption{\emph{Lower panel:} Recovered \ion{H}{1} Ly$\alpha$ pixel optical depth (black dots) as a function of redshift for the spectrum of Q1549+1933. The black dots at $\tau_{\rm HI}=10^{-5}$ are pixels with normalized flux greater than unity that have their values set to $\tau_{\rm min}$ (see the text for details). The points at $10^{4}$ show pixels that are saturated and that have insufficient available higher order lines for the recovery of their optical depths. The solid vertical line indicates the redshift of the QSO. The dashed vertical line, which is separated from the solid line by 3,000 $\rm km\,s^{-1}$, indicates the maximum redshift we consider for this spectrum. \emph{Upper panel:} Impact parameters of galaxies in this field (red stars) as a function of spectroscopic redshift.}\label{tauLBG}
\end{figure*}

\subsubsection{Physical properties}\label{DataGalProp}

These UV-selected star-forming galaxies have stellar masses  $\langle \rm log_{10} M_{*} /M_{\odot} \rangle =10.08\pm 0.51$ \citep{Shapley2005}. The typical star formation rates (SFRs) are $\sim 30$ $\rm M_{\odot}$ $\rm yr^{-1}$, where the SFRs of individual objects vary from $\approx7$ to $\approx200$ $\rm M_{\odot}$ $\rm  yr^{-1}$, and the mean SFR surface density is $\langle\Sigma_{\rm SFR}\rangle=2.9$ $\rm M_{\odot}$ $\rm yr^{-1}$ $\rm kpc^{-2}$ \citep{Erb2006a}. These stellar mass and SFR estimates assume a \citet{Chabrier2003} IMF. The galaxies show a correlation between their stellar mass and metallicity, but the relation is offset by 0.3 dex as compared to the local relation, with the same stellar mass galaxies having lower metallicity at $z\approx2.4$ \citep{Erb2006c}. Typical metallicities range from $\approx0.3\, Z_{\odot}$ for galaxies with $\langle M_{*} \rangle=2.7\times10^9 M_{\odot}$ to $\approx  Z_{\odot}$ for galaxies with $\langle M_{*} \rangle=1\times10^{11} M_{\odot}$.

As discussed in section~\ref{Redshifts}, ISM absorption lines are almost always blue-shifted with respect to the galaxy systemic redshift, and the Ly$\alpha$ emission line is always redshifted. 
These observed velocity offsets suggest that galaxy-scale outflows, with velocities of hundreds of $\rm km \, s^{-1}$, are the norm in these star-forming galaxies.  

\citet{Trainor2012} use the MultiDark simulation \citep{Prada2011}, together with a clustering analysis, to connect  galaxies from KBSS to dark matter halos. They find that this type of galaxy resides in  halos with masses above $10^{11.7}\, \rm M_{\odot}$, with a median halo mass of $\sim10^{11.9}\, \rm M_{\odot}$ \citep[similar conclusions were reached by e.g.][]{Adelberger2005b,Conroy2008}. The corresponding virial radii  are $\approx 75$ and $\approx87$  $\rm pkpc$, respectively, with circular velocities $\approx171\, \rm km\, s^{-1}$ and $\approx198\, \rm km\, s^{-1}$.

\subsection{QSO spectra}\label{QSOspectra}

The typical resolution of the QSO spectra is R~$\approx36,000$, and they were rebinned to pixels of $2.8\rm\, km\, s^{-1}$. The spectra were reduced using T. Barlow's MAKEE package and the continua were normalized using low-order spline fits. Further details about the QSO observations will be given in Steidel et al.\ (in preparation). The QSO redshift distribution can be seen in Figure~\ref{redshift} and a summary of the properties of the final  spectra is given in Table~\ref{tbl-1}. 

The redshift range that we consider when studying Ly$\alpha$ absorption in the spectrum of a QSO at redshift $z_{\rm QSO}$ is given by equation~(\ref{eq:zrange}).
The lower limit ensures that only Ly$\alpha$ redwards of the QSO's Ly$\beta$ emission line is considered, thus avoiding any confusion with the Ly$\beta$ forest from gas at higher redshifts. The upper limit is set to avoid contamination of the Ly$\alpha$ forest by material associated with the QSO and to avoid the QSO proximity effect. We verified that excluding 5,000 rather than 3,000 $\rm km \, s^{-1}$ gives nearly identical results.

The median S/N in the Ly$\alpha$ forest regions of the spectra ranges from 42 to 163 (Table \ref{tbl-1}).

A number of QSO spectra (5 out of 15) contain damped Lyman-$\alpha$ systems (DLAs) and sub-DLAs within the considered redshift range. We divided out the damping wings after fitting the Voigt profiles \citep[see][for more details on this procedure]{Rudie2012}. The saturated cores of the (sub-)DLAs were, however, flagged and excluded from the analysis, but this does not have a significant effect on our results.

\section{Pixel Optical Depth}\label{POD}

The goal of the pixel optical depth (POD) method is to use automatic algorithms (as opposed to manual fits) to find the best estimate of the optical depth of the \ion{H}{1} Ly$\alpha$ absorption line and those of various metal transitions as a function of redshift. Obtaining accurate pixel optical depths can be challenging due to the presence of noise, errors in the continuum fit, contamination and saturation. 

It is more instructive to plot optical depths than normalized fluxes because the optical depth is proportional to the column density of the absorbing species, while the flux is exponentially sensitive to this density. Hence, a power-law column density profile translates into a power-law optical depth profile and a lognormal column density distribution becomes a lognormal optical depth distribution. The shapes of the profiles and distributions common in nature are therefore preserved, which is not the case if we work with the flux.
Of course, if the spectral resolution is coarser than the typical line width, or if the wavelength coverage does not allow for the use of weaker components of multiplets to measure the optical depth of saturated lines, then the interpretation of the recovered optical depths is more complicated. Our spectral resolution is, however,  sufficient to resolve the Ly$\alpha$ lines and our wavelength coverage is sufficient to recover the optical depth of most saturated lines.

Another advantage of optical depths is that their dynamic range is unconstrained, whereas the normalized flux is confined to vary between zero and one. The large dynamic range does imply that one must use median rather than mean statistics, because the mean optical depth merely reflects the high optical depth tail of the distribution. While this problem could also be solved by using the mean logarithm of the optical depth, that would not resolve another issue resulting from the large intrinsic dynamic range: the observed optical depth distribution is cut off at the low absorption end by noise and at the high end by line saturation (and thus ultimately also by noise). However, errors in the tails of the distribution do not matter if we use median statistics (or other percentiles, provided they stay away from the noise).  

The POD statistics method was introduced by \citet{CowieSongaila1998} \citep[see also][]{Songaila1998}, and further improved by  \citet{Ellison2000}, \citet{Schaye2000b}, and \citet{Aguirre2002}. The POD method used  in this study is the one developed and tested by \citet{Aguirre2002}, and is identical to that of \citet{Ellison2000} for the case of Ly$\alpha$.

The \ion{H}{1} Ly$\alpha$ optical depth in each pixel is calculated from the normalized flux $F(\lambda)$,
\begin{equation}
\tau_{\rm Ly\alpha}(\lambda)=-\ln (F(\lambda)).
\end{equation}
Pixels with $\tau_{\rm Ly\alpha}<0$ (which can occur in the presence of noise) are assigned an optical depth of $\tau_{\rm min}=10^{-5}$, which is smaller than any recovered optical depth.  

Pixels that have $F(\lambda) \le N_{\sigma}\sigma_{\lambda}$ are considered saturated, where $\sigma_{\lambda}$ is the rms noise amplitude at the given pixel and $N_{\sigma}$ is a parameter that we set to 3 (see Aguirre, Schaye \& Theuns, 2002 for more details). One advantage of the POD method is that it is easy to recover a good estimate of the Ly$\alpha $ optical depth in these saturated pixels by using the available higher order Lyman lines. The recovered Ly$\alpha$ optical depth is then given by
\begin{equation}
\tau_{\rm Ly\alpha}^{\rm rec}=\rm min\{ \tau_{\rm Ly \it n}\it f_{\rm Ly\alpha}\lambda_{\rm Ly\alpha}\slash \it f_{\rm Ly \it n}\lambda_{\rm Ly {\it n}}\},
\label{eq:HOeq}
\end{equation}
where $ f_{\rm Ly \it n}$ is the oscillator strength of the $n$th order Lyman line and  $\lambda_{\rm Ly \it n}$ is its rest wavelength ($n=1$ corresponds to $\alpha$, $n=2$ to $\beta$, etc.). Taking the minimum optical depth (Equation ~\ref{eq:HOeq}) minimizes the effect of contamination by other lines. Higher order lines used for Ly$\alpha$ optical depth recovery are those that lie in the wavelength range of the spectrum, and for which $N_{\sigma}\sigma_{n}\le F(\lambda_{\rm Ly \it n})\le 1-N_{\sigma}\sigma_{n}$, where $\sigma_{n}$ is the noise at $\lambda_{\rm Ly \it n}$. If none of the available higher order lines satisfy this criterion, or if none of the higher order lines are available, then the pixel is set to a high optical depth (we use $\tau_{\rm max}=10^{4}$). An example of the recovered optical depth in one QSO spectrum is shown in Figure~\ref{tauLBG}, together with the positions of galaxies in the same field. 

Setting pixels to $\tau_{\rm min}$ and  $\tau_{\rm max}$ allows them to correctly influence the median and other percentiles. The actual values they are set to are unimportant as long as $\tau_{\rm min}<\tau^{\rm rec}_{\rm Ly\alpha}<\tau_{\rm max}$ for the percentile of interest. In other words, as long as the $\tau_{\rm min}$ or  $\tau_{\rm max}$ are on the same side of a given percentile as the true optical depth that they are replacing, their actual values are unimportant. 

We use the POD method because: a) It is fast, robust, and automatic, which means it can deal with large amounts of data, both observed and simulated; b) It makes it straightforward to exploit the full dynamic range measurable from our spectra; c) The fact that the optical depth is directly proportional to the column density of neutral hydrogen makes it easy to interpret the results (see \S\ref{Interpretation}). While the POD method has clear advantages, it is important to note that it does not replace the more traditional method of decomposing the absorption spectra into Voigt profile components. For example, in its simplest form, the POD method does not make direct use of information about the line widths, which are clearly of interest as they contain important information about the temperature and the small-scale velocity structure of the absorbing gas.  An analysis of the KBSS data based on Voigt profile decompositions is presented in \citet{Rudie2012}.

\section{Ly$\alpha$ absorption near galaxies}\label{Results}

After recovering the optical depths in each pixel of a QSO spectrum, we can build a map of the galaxies'  average surroundings, as a function of the transverse distance from the line of sight (LOS), i.e.\  the impact parameter, and the velocity separation from the galaxy along the LOS. We sometimes convert velocity separation into distance by assuming that it is due to the Hubble flow. However, as we will show in sections~\ref{2Ddistribution} and \ref{sec:distortions}, this is  a poor approximation  at small velocity differences from galaxies. We will refer to distances computed under the assumption of zero peculiar velocities (and zero redshift errors) as ``Hubble distances".

Different galaxies and QSO fields are combined, meaning that the median POD for a given impact parameter and velocity difference is estimated over all galaxies irrespective of which field they come from, without applying any weighting. Each galaxy therefore provides an array of pixels with varying velocity difference but fixed impact parameter.

For reference, we note that 0.2, 2.1, and 9.9 percent of the pixels in our sample are at 3-D Hubble distances smaller than 200, 500, and 1000 pkpc, respectively. 

\subsection{2-D Maps of the median Ly$\alpha$ absorption} \label{2Ddistribution}
\begin{figure*}
\centering
\resizebox{0.3\textwidth}{!}{\includegraphics*[85,370][135,470]{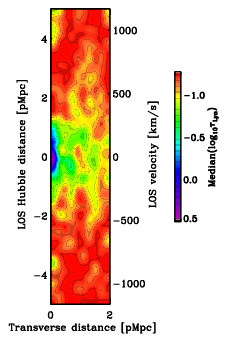}}
\quad\quad
\resizebox{0.396\textwidth}{!}{\includegraphics*[85,370][151,470]{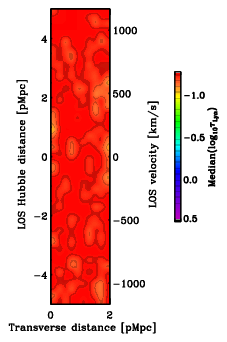}}
\caption{\emph{Left:} Median \ion{H}{1} Ly$\alpha$ absorption as a
function of transverse and LOS distance from $\langle
z\rangle\approx2.4$ star-forming galaxies. The bin size is 200~pkpc and
the images have been smoothed by a 2-D Gaussian with FWHM equal to the
bin size. The lower limit of the color scale corresponds to the median
optical depth of all pixels (log$_{10}=-1.27$), while the upper limit is set to log$_{10}=0.5$ (very close to the maximum optical depth in this map, log$_{10}=0.47$). Absorption is clearly enhanced close to
galaxies, out to at least 2~pMpc in the transverse direction,
but only out to $\approx1.5$~pMpc along the LOS. This anisotropy suggests
large-scale infall of gas. On the other hand, on small scales
the absorption declines more rapidly in the transverse direction than
in the LOS direction. \emph{Right:} Results after randomizing the
galaxy redshifts while keeping their impact parameters fixed. The fact
that the correlation does not vary systematically with distance
indicates that the features in the left panel are
real.}\label{2DdifferentialFixed}
\end{figure*}

\begin{figure}
\resizebox{0.45\textwidth}{!}{\includegraphics*[75,383][172,454]{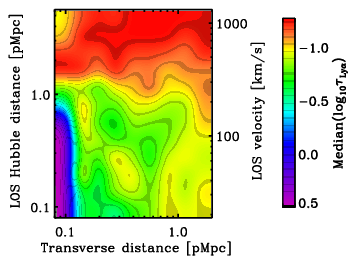}}

\caption{Similar to Fig.~\protect\ref{2DdifferentialFixed}, but we
do not separate positive and negative velocity differences between
absorbers and galaxies and we use logarithmic bins and axes (bins are
separated by 0.15 dex). The map is smoothed by a 2-D Gaussian with
FWHM equal to the bin size. The color scale is saturated at the high
optical depth end (at the smallest impact parameters the optical
depth is log$_{10}\approx1.15$). }
\label{2DdifferentialLOG}
\end{figure}

The left panel of Figure~\ref{2DdifferentialFixed}  shows the logarithm of the median  $\tau_{\rm Ly\alpha}$  as a function of  the transverse and the LOS separations from galaxies. The distance bin size in this plot is 200 by 200 pkpc (which corresponds to $\approx 46 ~\rm km\, s^{-1}$ along the LOS), and the image has been smoothed with a 2-D Gaussian with FWHM equal to the bin size. Note that we take a galaxy-centered approach in constructing this image. Each galaxy contributes a column of pixels whose position along the $x$-axis corresponds to the galaxy's impact parameter. A single pixel can be used multiple times: once for each galaxy whose separation from the pixel falls within the range plotted. This is the first published 2-D absorption map around galaxies. Studies by e.g. \citet{Adelberger2003, Ryan-Weber2006, Wilman2007} and \citet{Shone2010} show cross-correlation maps of galaxies with Ly$\alpha$ absorption systems, which is not equivalent. The map shows a strong correlation between the  Ly$\alpha$ absorption strength and the distance to the galaxies. 

The right panel of Figure~\ref{2DdifferentialFixed} shows the same
data, after randomizing the galaxy redshifts. We randomize the galaxy
redshifts within the Ly$\alpha$ forest region of each QSO spectrum
while keeping their impact parameters unchanged in order to preserve the number of pixels per galaxy. In this way we can estimate the magnitude
of the fluctuations in the absence of correlations
between the locations of galaxies and absorbers. We can see that the
signal is lost, which implies that the features seen in the left panel
are real. Finally, we note that because a single galaxy contributes a full column of pixels at its impact parameter, bins along the LOS are somewhat correlated. On the other hand, bins in the transverse
direction are independent. In Appendix~\ref{RankCorrelation} we demonstrate that along the LOS the errors are significantly correlated for scales $\lesssim 10^2~\rm km\,s^{-1}$.

In Figure~\ref{2DdifferentialFixed} we kept the positive and negative
velocity differences between absorbers and galaxies  separated, which
gives insight into the amount of noise and sample variance. However,
given that the Universe is statistically isotropic\footnote{The Universe is denser at higher redshift and therefore we expect more absorption at positive velocity differences in Figure~\ref{2DdifferentialFixed} than at negative velocities, but this effect is negligible on the scales that we consider. The mean transmission varies by $\approx2\times10^{-3}$ over $10^3\,\rm km\, s^{-1}$ at $z = 2.4$ \citep[e.g.][]{Schaye2003}, which is much smaller than the excess absorption in the vicinity of galaxies.} (and our
observations are consistent with this assumption), we can increase the
S/N by considering only absolute velocity differences.
Figure~\ref{2DdifferentialLOG} shows a map of gas around galaxies
by taking into account only absolute velocity separations. We use logarithmic axes, which brings out the small-scale anisotropy more clearly. 

The most prominent feature is the region of strongly enhanced absorption that extends to $\sim 10^2$ pkpc in the transverse direction, but to $\sim 1\,$pMpc ($\sim 200\, \rm km\, s^{-1}$) along the LOS. This redshift space distortion (called the ``finger of God'' effect in galaxy redshift surveys) could be due to redshift errors and/or peculiar velocities. A more subtle anisotropy is visible on large scales. In the transverse direction the correlation between absorption strength and galaxy position persists out to the maximum impact parameter we consider, 2~pMpc, whereas it becomes very weak beyond $\approx 1.5\,$pMpc ($\approx 300\, \rm km\, s^{-1}$) along the LOS, as is most clearly visible in Figure~\ref{2DdifferentialFixed}. If real, such a feature would imply infall of gas on large scales, i.e.\ the \citet{Kaiser1987} effect. Other scenarios are unlikely, as the signal distortion along the LOS can only be caused by redshift errors and peculiar velocities. However, redshift errors would act to spread the signal along the LOS rather than compressing it, which leaves peculiar velocities as the only other possibility. We will examine the significance of these anisotropies in the next section.

\subsection{Redshift space distortions}
\label{sec:distortions}

\begin{figure*}
\centering
\resizebox{0.9\textwidth}{!}{\includegraphics*[105,360][475,685]{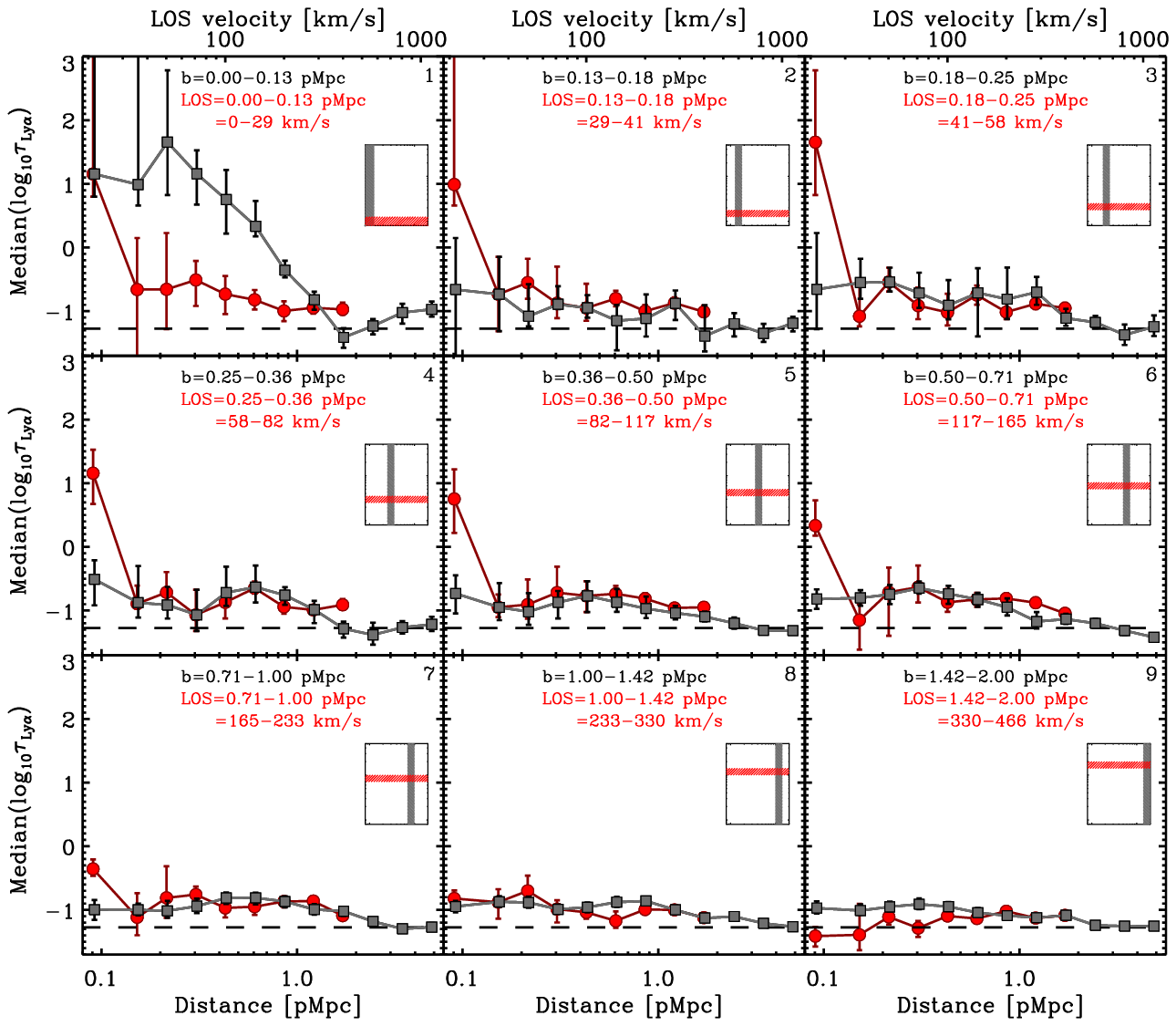}}
\caption{Cuts through the unsmoothed 2-D map from Fig.~\ref{2DdifferentialLOG}, where red circles show median profiles in the transverse direction (i.e.\ parallel to the $x$-axis of Fig.~\ref{2DdifferentialLOG}), and grey squares show cuts along the LOS (i.e.\ parallel to the $y$-axis of Fig.~\ref{2DdifferentialLOG}). The range of LOS separations and impact parameters included in the strips along the $y$- resp.\ $x$-axis of Fig.~\ref{2DdifferentialLOG} is indicated in each panel. Distances along the LOS have been computed from the corresponding velocity interval under the assumption of pure Hubble flow. Hence, differences between the red and black curves indicate redshift space distortions. The dashed horizontal line shows the median \lya\ optical depth of all the pixels in the spectra. With the exception of the last red circles and the corresponding grey squares, which correspond to 1.42 -- 2.00~pMpc, the differences between the red and black curves in the first 8 panels indicate that the signal has been smoothed in the LOS direction, probably as a result of redshift errors and, most importantly, small-scale peculiar velocities. On the other hand, the differences revealed by the most distant points in the first six panels or, equivalently (see text), by the first six points in the last panel, imply compression along the LOS. This compression indicates the presence of large-scale infall onto the galaxies. Data plotted in this figure are tabulated in Table~\ref{tbl-2}.}\label{2Dcuts}
\end{figure*}

\begin{deluxetable*}{llllllllll}
\tabletypesize{\scriptsize}
\tablecaption{Data from Figure~\ref{2DdifferentialLOG}: the median log$_{10}\tau_{\rm Ly\alpha}$ as a function of impact parameter (rows) and distance along the LOS (columns), with the 1$\sigma$ confidence interval. 
\label{tbl-2}}
\tablewidth{0pt}
\tablehead{\colhead{Distance [pMpc]}&	\colhead{0.00-0.13}	& 	\colhead{0.13-0.18}	& \colhead{0.18-0.25}	& \colhead{0.25-0.36}	& \colhead{0.36-0.50}	& \colhead{0.50-0.71}	& \colhead{0.71-1.00}	&\colhead{1.00-1.42}	& \colhead{1.42-2.00}}
\startdata
0.00-0.13& 1.15$^{+2.85}_{-0.35}$& -0.66$^{+0.81}_{-1.05}$& -0.66$^{+0.88}_{-0.62}$& -0.51$^{+0.30}_{-0.41}$& -0.73$^{+0.29}_{-0.32}$& -0.82$^{+0.15}_{-0.15}$& -1.00$^{+0.15}_{-0.16}$&-0.95$^{+0.09}_{-0.09}$& -0.97$^{+0.11}_{-0.08}$\\
0.13-0.18& 0.99$^{+3.01}_{-0.33}$& -0.73$^{+0.59}_{-0.58}$& -0.55$^{+0.37}_{-0.25}$& -0.88$^{+0.58}_{-0.23}$& -0.95$^{+0.38}_{-0.20}$& -0.80$^{+0.12}_{-0.12}$& -1.00$^{+0.10}_{-0.09}$&-0.87$^{+0.09}_{-0.08}$& -1.01$^{+0.10}_{-0.06}$\\
0.18-0.25& 1.65$^{+1.13}_{-0.83}$& -1.08$^{+0.50}_{-0.16}$& -0.54$^{+0.23}_{-0.15}$& -0.91$^{+0.29}_{-0.21}$& -1.02$^{+0.29}_{-0.21}$& -0.75$^{+0.15}_{-0.15}$& -1.01$^{+0.16}_{-0.11}$&-0.88$^{+0.09}_{-0.08}$& -0.95$^{+0.08}_{-0.07}$\\
0.25-0.36& 1.16$^{+0.37}_{-0.48}$& -0.89$^{+0.28}_{-0.07}$& -0.71$^{+0.32}_{-0.23}$& -1.07$^{+0.40}_{-0.25}$& -0.87$^{+0.18}_{-0.25}$& -0.65$^{+0.11}_{-0.09}$& -0.94$^{+0.12}_{-0.11}$&-0.99$^{+0.07}_{-0.07}$& -0.91$^{+0.09}_{-0.08}$\\
0.36-0.50& 0.75$^{+0.47}_{-0.53}$& -0.95$^{+0.20}_{-0.15}$& -0.91$^{+0.39}_{-0.22}$& -0.72$^{+0.41}_{-0.21}$& -0.77$^{+0.23}_{-0.26}$& -0.74$^{+0.12}_{-0.16}$& -0.81$^{+0.09}_{-0.09}$&-0.96$^{+0.07}_{-0.08}$& -0.95$^{+0.07}_{-0.07}$\\
0.50-0.71& 0.33$^{+0.40}_{-0.16}$& -1.15$^{+0.25}_{-0.47}$& -0.71$^{+0.39}_{-0.68}$& -0.63$^{+0.34}_{-0.24}$& -0.87$^{+0.21}_{-0.15}$& -0.83$^{+0.10}_{-0.12}$& -0.81$^{+0.08}_{-0.10}$&-0.88$^{+0.08}_{-0.08}$& -1.04$^{+0.07}_{-0.07}$\\
0.71-1.00& -0.36$^{+0.15}_{-0.11}$& -1.11$^{+0.38}_{-0.28}$& -0.81$^{+0.50}_{-0.31}$& -0.76$^{+0.13}_{-0.15}$& -0.97$^{+0.19}_{-0.15}$& -0.95$^{+0.14}_{-0.13}$& -0.87$^{+0.08}_{-0.09}$&-0.86$^{+0.06}_{-0.06}$& -1.09$^{+0.07}_{-0.05}$\\
1.00-1.42& -0.82$^{+0.13}_{-0.16}$& -0.88$^{+0.20}_{-0.26}$& -0.70$^{+0.24}_{-0.20}$& -0.99$^{+0.14}_{-0.21}$& -1.04$^{+0.11}_{-0.10}$& -1.17$^{+0.15}_{-0.11}$& -0.99$^{+0.08}_{-0.08}$&-1.00$^{+0.05}_{-0.05}$& -1.12$^{+0.06}_{-0.04}$\\
1.42-2.00& -1.41$^{+0.14}_{-0.16}$& -1.39$^{+0.49}_{-0.24}$& -1.11$^{+0.15}_{-0.12}$& -1.28$^{+0.11}_{-0.14}$& -1.10$^{+0.08}_{-0.07}$& -1.13$^{+0.08}_{-0.08}$& -1.02$^{+0.07}_{-0.07}$&-1.13$^{+0.06}_{-0.06}$& -1.09$^{+0.05}_{-0.05}$\\
2.00-2.83& -1.23$^{+0.11}_{-0.13}$& -1.20$^{+0.17}_{-0.20}$& -1.17$^{+0.10}_{-0.09}$& -1.38$^{+0.19}_{-0.15}$& -1.20$^{+0.09}_{-0.09}$& -1.20$^{+0.07}_{-0.07}$& -1.18$^{+0.05}_{-0.06}$&-1.10$^{+0.04}_{-0.04}$& -1.24$^{+0.04}_{-0.04}$\\
2.83-3.99& -1.02$^{+0.14}_{-0.17}$& -1.35$^{+0.15}_{-0.13}$& -1.37$^{+0.16}_{-0.16}$& -1.27$^{+0.10}_{-0.09}$& -1.31$^{+0.06}_{-0.05}$& -1.31$^{+0.06}_{-0.07}$& -1.30$^{+0.05}_{-0.05}$&-1.21$^{+0.05}_{-0.04}$& -1.25$^{+0.04}_{-0.04}$\\
3.99-5.64& -0.97$^{+0.12}_{-0.11}$& -1.19$^{+0.10}_{-0.13}$& -1.24$^{+0.18}_{-0.15}$& -1.22$^{+0.12}_{-0.11}$& -1.31$^{+0.06}_{-0.05}$& -1.42$^{+0.05}_{-0.06}$& -1.27$^{+0.06}_{-0.06}$&-1.26$^{+0.03}_{-0.03}$& -1.25$^{+0.03}_{-0.03}$\\
\enddata
\end{deluxetable*}

Figure~\ref{2Dcuts} shows ``cuts" along the first 9 rows and columns of the 2-D map shown in Figure~\ref{2DdifferentialLOG}, spanning 0--2 pMpc. The red circles and grey squares show the profiles along the transverse and LOS, respectively. As indicated in the insets, these cuts correspond, respectively, to the horizontal and vertical strips in Fig.~\ref{2DdifferentialLOG}.
Observe that the transverse and LOS directions are identical for the
$n^{\rm th}$ data point in the $n^{\rm th}$ panel, which corresponds
to the intersection of the horizontal and vertical strips shown in the
insets. In other words, where the horizontal and vertical strips meet, the data point is replicated as both a red and a black symbol. Note also that the $n^{\rm th}$ red circle (black square) of the $m^{\rm th}$ panel is identical to the $m^{\rm th}$ black square (red circle) of the $n^{\rm th}$ panel. The figure thus contains redundant information. For example, the LOS direction (i.e.\ black squares) in the $1^{\rm st}$ panel shows all the $1^{\rm st}$ data points appearing in the transverse direction (i.e.\ red circles) of the other panels. Similarly, the $2^{\rm nd}$ panel shows all the $2^{\rm nd}$ transverse data points, etc.

As LOS separations have been computed under the assumption of pure Hubble flow, any significant difference between the red and black curves must be due to redshift space distortions (assuming the Universe is isotropic in a statistical sense). By comparing the two curves in each panel, we can therefore identify redshift space distortions and assess their significance.

The error bars in Figure~\ref{2Dcuts}, as well as those in subsequent plots, were computed by bootstrapping the galaxy sample 1,000 times. That is, for each bootstrap realization we randomly select galaxies,  where each galaxy could be selected multiple times, until the number of galaxies from the original set is reached. We calculate the results for each realization and the errors show the 1$\sigma$ confidence interval. As demonstrated in Appendix~\ref{RankCorrelation}, along the LOS the errors are correlated for separations $\lesssim 10^2~\rm km\, s^{-1}$ (i.e.\ black squares in each panel are correlated on these scales), but not in the transverse direction (i.e.\ red circles in each panel are independent).

The number of galaxies per transverse bin depends slightly on the velocity difference, because galaxies separated by $\Delta v ~\rm km\, s^{-1}$ from the \lya\ forest region (as defined in section~\ref{QSOspectra}) will still contribute pixels to bins with velocity separations greater than $\Delta v$. From small to large impact parameters, the number of galaxies contributing to the $1^{\rm st}$ ($9^{\rm th}$) velocity bin is 14 (16), 8 (8), 11 (12), 22 (23), 47 (48), 58 (62), 96 (99), 175 (180), 202 (211). Thus, the black squares in the first panel, as well as the first red circle in all panels, are based on 14 -- 16 galaxies. Similarly, the black squares in the $9^{\rm th}$ panel, as well as the last red circle in all panels, are based on 202 -- 211 galaxies. 

The first panel, which shows cuts near the axes of Figure~\ref{2DdifferentialLOG}, clearly shows that out to $\sim 1$ pMpc, which translates into $\lesssim 200\, \rm km\, s^{-1}$ along the LOS (see the top $x$-axis), the absorption is stronger along the LOS (black squares) than in the transverse direction (red circles). In panels 2 -- 7, which correspond to strips offset by 0.13 -- 1.00 pMpc from the axes in Figure~\ref{2DdifferentialLOG}, the smearing in the LOS direction manifests itself as enhanced absorption at small impact parameters relative to the signal at small LOS separations. The confidence level associated with the detected discrepancy between the two directions, which we estimate as the fraction of bootstrap realizations in which the sign of the discrepancy is reversed, is at least 99\% for each data point out to 1~pMpc (i.e.\ points 2--7 of the first panel). 

The differences on scales $\lesssim 1$ pMpc ($\lesssim 233 \rm km\, s^{-1}$) can be explained by two effects. Firstly, we expect gas in and around galaxy halos to have peculiar velocities comparable to the circular velocity of the halos, i.e.\ $\simeq 200\rm\,km\, s^{-1}$  (and possibly significantly higher for outflowing gas). Secondly, as discussed in Section~\ref{Redshifts}, there are random errors in the redshift measurements of $\approx 130\, \rm km\, s^{-1}$ for LRIS redshifts,  and $\approx 60 \rm\, km\, s^{-1}$ for the NIRSPEC subsample. These two effects smooth the signal in velocity space on the scale that is a result of the combination of these velocities. In Appendix~\ref{sec:zerrors} we show that redshift errors may be able to account for the observed smoothing. However, this does not prove the observed elongation along the LOS is due to redshift errors rather than peculiar velocities. In fact, since the elongation is more extended than the typical redshift errors ($\sim 200 \rm\,km\, s^{-1}$ vs.\ $\approx 130\, \rm km\, s^{-1}$), but similar to the expected circular velocities ($\approx 200 \, \rm km\, s^{-1}$), it is likely that the finger of God effect is dominated by small-scale peculiar velocity gradients due to virial motions, infall, and/or outflows. This would not be surprising considering the observations of large velocity fields in the gas in star-forming galaxies \citep[e.g.][]{Steidel2010}. \citet{Rudie2012} use the KBSS sample to study Ly$\alpha$ absorption near galaxies by decomposing the forest lines into Voigt profiles, and they also detect the finger of God effect at small impact parameters. After presenting a more detailed analysis of the impact of redshift errors, they conclude that  redshift errors alone  cannot account for the detected elongation of the absorption signal along the LOS, which strengthens our conclusion. It is also worth noting that even if redshift errors were solely responsible for the finger of God effect, we can still safely conclude that the detected peculiar motions of Ly$\alpha$ absorbing gas near galaxies do not exceed $\sim200\rm\, km\, s^{-1}$. However, this does not mean that such gas does not reach higher velocities very close to galaxies, as the smallest impact parameter in the KBSS sample is $\approx55$ pkpc.

On the other hand, at distances $> 1.42$~pMpc the situation is reversed: the absorption is compressed in the LOS direction. For impact parameters 1.42 -- 2~pMpc and LOS separations 0-- 0.71~pMpc (0 -- 165~$\rm km \, s^{-1}$; $9^{\rm th}$ (i.e.\ last) red circles in panels 1--6) the absorption is stronger than for impact parameters 0-- 0.71~pMpc and LOS separations 1.42 -- 2~pMpc ($9^{\rm th}$ black square in panels 1--6). The same information is collected in the last panel, which corresponds to strips separated by 1.42 -- 2.00 pMpc from the axes. The first six black squares are higher than the corresponding red circles, which implies that the absorption is enhanced along the LOS relative to the direction transverse to the LOS. Observe that this enhancement is absent from all other panels. The confidence level with which this discrepancy is detected is, from small to large scales, 99.4\%, 77.9\%, 83.2\%, $99.8$\%, 91.6\% and 82.6\% for points 1--6, respectively. As mentioned above, these confidence levels represent the fraction of bootstrap realizations in which the sign of the discrepancy is the same as for the original sample. 

In order to estimate the significance of the compression along the LOS, we combine different measurements of the difference between pairs of points. For each pair of points \emph{i} (points 1--6 in the last panel of Figure~\ref{2Dcuts}), we first compute the difference between median optical depths, $d_{i}=a_{i}-b_{i}$, where $a_{i}$ and $b_{i}$ are the observed values, and the associated error $s_{i}=\sqrt{s_{a,i}^2+s_{b,i}^2}$, where $s_{a,i}$ and $s_{b,i}$ are errors on measurements $a_{i}$ and $b_{i}$. After that we compute the weighted average difference for the 6 pairs of points, $\langle d\rangle = \sum_{i=1}^{i=6}(d_{i}/s_{i}^2)/\sum_{i=1}^{i=6}(1/s_{i}^2)$, and the corresponding error, $s^2 = 1 / \sum_{i=1}^{i=6}(1/s_{i}^2)$. The significance is then $\langle d\rangle/s$ (assuming Gaussian errors). This procedure suggests that the compression along the LOS is detected at 3.5$\sigma$.

This compression along the LOS can be explained by large-scale infall, i.e.\ the Kaiser effect. As the absorbing
gas must be cool, i.e.\ $T \sim 10^4~\rm K \ll T_{\rm vir}$, to be
visible in HI, this form of gas accretion could be called ``cold
accretion'' \citep[e.g.][]{Keres2005}, although we note that this term is most often used in the context of cold streams within the virial radii of halos hosting galaxies. We conclude that we have an unambiguous and highly significant detection of cool gas falling towards star-forming galaxies at $z \approx 2.4$. The large-scale infall of gas onto haloes and its impact on the absorption signal has been a topic of several theoretical studies \citep[e.g.][]{Kollmeier2003, Barkana2004, KimCroft2008}, and is also investigated in Rakic et al. (in preparation) where we will compare the observational results with cosmological hydrodynamical simulations.

The dashed, horizontal lines in Figure~\ref{2Dcuts} indicate the median \lya\ optical depth of all pixels in the \lya\ forest regions of the QSO spectra. In the transverse direction we do not probe sufficiently large distances to see the signal disappear: for all but the last panel the red curves stay above the dashed lines out to impact parameters of 2~pMpc. In the LOS direction (black curves) we do see convergence for separations $\gtrsim 3$~pMpc. The last black square in the first panel is an outlier, but note that it is based on only 16 galaxies, whereas the same points in panels 5-9 are based on 49--217 galaxies. Indeed, according to the redshift randomization method described in the next section, the last black square in the first panel is only a 1.8$\sigma$ outlier, whereas the last red circles of panels 1-6 (or, equivalently, the first 6 black squares in the last panel) represent detections of excess absorption with significance varying between 2.6$\sigma$ and $> 3\sigma$. 

The median log($_{10}\tau_{\rm Ly\alpha}$) and 1$\sigma$ confidence intervals from  Figure~\ref{2DdifferentialLOG} are tabulated in Table~\ref{tbl-2}.

\subsection{Ly$\alpha$ absorption as a function of 3-D Hubble distance}

\begin{figure}
\epsscale{1.1}
\plotone{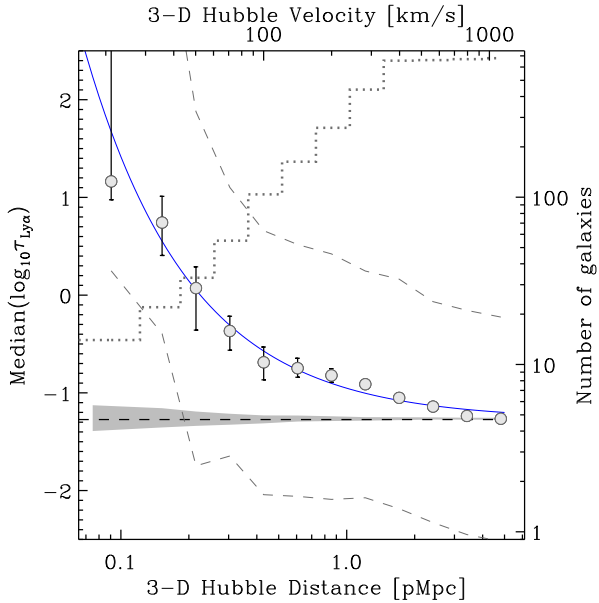}
\caption{Median log$_{10}\tau_{\rm Ly\alpha}$ as a function of 3-D Hubble distance from galaxies. Distance bins are separated by 0.15 dex. The dashed lines show the 15.9\% and 84.1\% percentiles, i.e.\ the $1\sigma$ scatter of the PODs around the median.  Note that the errors are correlated over scales $\lesssim 10^2 ~\rm km \, s^{-1}$ (see Appendix~\ref{RankCorrelation}). The horizontal dashed line shows the median of all pixels in the spectra. The grey shaded region shows the 1$\sigma$ detection threshold. It shows the 1$\sigma$ confidence interval for the median(log$_{10}\tau_{\rm Ly\alpha}$) that we obtain after randomizing the redshifts. The dotted line shows the number of galaxies in each distance bin (right $y$-axis). The blue curve shows the best-fit power-law: log$_{10}\tau_{\rm Ly\alpha}=0.32d_{\rm 3D}^{-0.92}-1.27$, where $d_{\rm 3D}$ is the 3-D Hubble distance. }\label{Hubble}
\end{figure}

Figure~\ref{Hubble} shows the median $ \rm Ly\alpha$ optical depth in radial bins around the galaxy positions, where we assumed that velocity differences between absorbers and galaxies are entirely due to the Hubble flow. We emphasize that Figures~\ref{2DdifferentialLOG} and \ref{2Dcuts} show that this is not a good approximation, particularly for distances $\lesssim 1\,$pMpc. It does, however, provide us with a compact way to present a lot of information.

The 3-D Hubble distance, as we call it, is therefore just $d_{\rm 3D}=\sqrt{b^2+(H(z)\Delta v)^2}$, where $b$ is the galaxy's impact parameter, $H(z)$ is the Hubble parameter, and $\Delta v$ is the velocity separation between an absorber and the galaxy. As mentioned in Section~\ref{galaxies}, we use only galaxies with impact parameters smaller than 2 pMpc, even when making Figure~\ref{Hubble}. Hence, distances $\gtrsim 2$ pMpc reflect mostly LOS separations.

The horizontal dashed line shows the median level of absorption in the Ly$\alpha$ forest pixels. The significance of the excess absorption can be estimated by comparing the error bars, which indicate the $1\sigma$ confidence intervals determined by bootstrap resampling the galaxies, to the difference between the data points and the horizontal dashed line. More precisely, we can estimate the confidence level  associated with the detection of excess absorption as $1-2f_{\rm b,low}$, where $f_{\rm b,low}$ is the fraction of 1,000 bootstrap realizations for which the data point falls below the dashed line. This method indicates that within 2.8 pMpc excess absorption is detected with greater than 99.7\% significance (i.e.\ $>3\sigma$). For 2.8 -- 4.0 pMpc the significance is 87\% (i.e.\ $1.5\sigma$), while the absorption is consistent with random beyond 4~pMpc.

The dotted curve and the right $y$-axis show the number of galaxies contributing to each bin. Since the inner few bins contain only a few tens of galaxies each (14 for the first bin), the bootstrap errors may not be reliable for these bins. The significance of the excess of absorption can be estimated more robustly by making use of the fact that each QSO spectrum provides many independent spectral regions at the impact parameter of each galaxy. We can do this by comparing the excess absorption to the grey region, which indicates the $1\sigma$ detection threshold and which was determined by re-measuring the median $\rm log_{10}\tau_{\rm Ly\alpha}$ after randomizing the galaxy redshifts (while keeping the impact parameters fixed). The grey shaded region shows the $1\sigma$ confidence interval obtained after doing this 1,000 times. For each distance bin, the confidence level of the detection is then given by $1-2f_{z,{\rm high}}$, where $f_{z,{\rm high}}$ is the fraction of realizations resulting in a median optical depth that is higher than actually observed. In agreement with the errors estimated by bootstrap resampling the galaxies, we find that the significance is $>99.7$\% within 2.8~pMpc and that there is no evidence for excess absorption beyond 4~pMpc. For 2.8--4.0 pMpc the significance of the detection is, however, larger than before: 99.2\% (i.e.\ $2.7\sigma$). We conclude that the absorption is significantly enhanced out to at least 3~pMpc proper, which is $7h^{-1}$~cMpc.

The fact that the absorption is enhanced out to several pMpc is in good agreement with \citet{Adelberger2005}, who measured the mean flux as a function of 3-D Hubble distance. However, the profile measured by \citet{Adelberger2005} is much flatter. Converting their data into optical depths, they measure $\log_{10}\tau_{\rm Ly\alpha} \approx -0.1$ in their innermost bin, which extends to about 200~pkpc. This is about an order of magnitude lower than our median recovered optical depth at this distance. Conversely, at large distances their mean flux asymptotes to 0.765, or  $\log_{10}\tau_{\rm Ly\alpha} \approx -0.57$, which is much higher than our asymptotic median optical depth of $\log_{10}\tau_{\rm Ly\alpha} \approx -1.27$, even though we measure a similar mean flux of 0.806, or $\log_{10}\tau_{\rm Ly\alpha} \approx -0.67$. Thus, our dynamic range is about two orders of magnitude larger than that of \citet{Adelberger2005}. This difference arises because we use median optical depth rather than mean flux statistics and because we use higher order Lyman lines to recover the optical depth in saturated lines.

The dashed curves show the 15.9\% and 84.1\% percentiles, indicating the $1\sigma$ scatter in the PODs (which is obviously much larger than the error in the median). It is important to note that, except on the smallest scales ($\lesssim 200$ pkpc), the scatter is similar to or larger than the median excess absorption. Hence, there will be a wide range of PODs for all separations probed here.

Finally, the blue line shows the best-fit power-law through the data points,
\begin{equation}
{\rm median}(\log_{10}\tau_{\rm Ly\alpha}) = (0.32 \pm 0.08) d_{\rm 3D}^{-0.92 \pm 0.17} - 1.27,
\end{equation}
where we required the fit to asymptote to the median of all pixels (horizontal, dashed line in Figure~\ref{Hubble}). 

\begin{figure*}
\epsscale{1.1}
\plotone{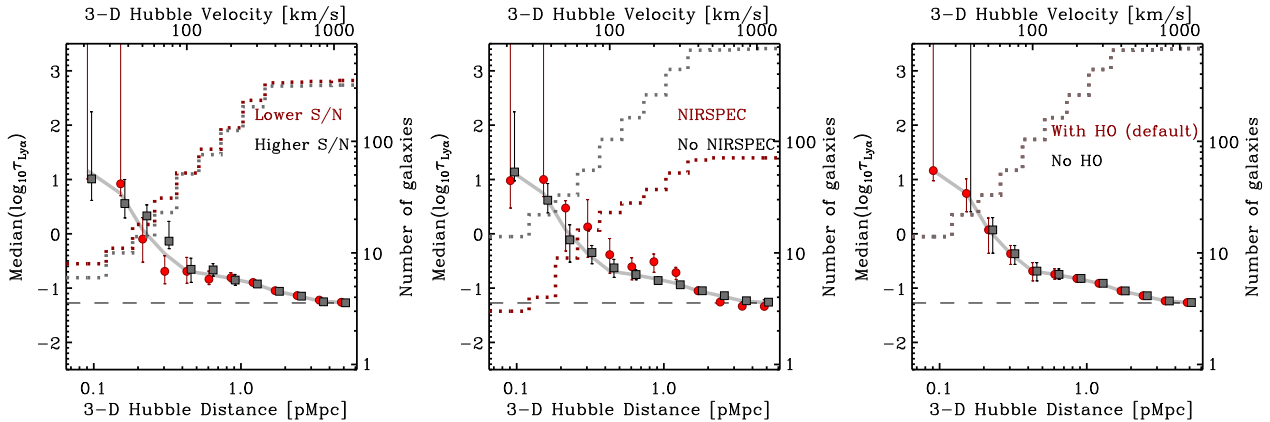}
\caption{Similar to Figure~\ref{Hubble} but: \emph{the left panel} shows the differences arising when using only higher or lower S/N QSO spectra; \emph{the middle panel} shows the results for the galaxy subsample with near-IR redshifts, and for the whole sample when not using redshifts measured from near-IR lines; \emph{the right panel} shows the effect of relying solely on Ly$\alpha$ and neglecting higher order transitions, which otherwise allow the recovery of the optical depth in saturated Ly$\alpha$ pixels. In all panels the black points have been slightly offset for clarity and the grey curve shows the result for the default sample and method.}\label{HubbleVariations}
\end{figure*}

\subsubsection{Testing the robustness}

In this section we will show that the results we presented (2-D and 3-D absorption profiles) are robust to changes in the S/N ratio of the QSO spectra, to the omission of NIRSPEC or non-NIRSPEC redshifts, to the exact redshift calibration, and, for optical depths $\ll 10$, to the use of higher order Lyman lines. We have chosen to demonstrate this using the plots of median absorption versus 3-D Hubble distance, because these offer a compact summary of the data. 

In the left panel of Figure~\ref{HubbleVariations} we compare the median Ly$\alpha$ absorption as a function of 3-D Hubble distance for the lower and higher  S/N subsamples of the QSO spectra (with median S/N ratios of $\approx65$ and $\approx85$ respectively). It appears that better data yields slightly more absorption at 3-D Hubble distances of $\approx 0.3$~pMpc, but the differences are not significant. 
 
The middle panel of Figure~\ref{HubbleVariations} compares the subset of 71 galaxies for which redshifts have been measured from nebular emission lines using the NIRSPEC instrument (red circles) with the default sample (grey curve) as well as with the result obtained when we ignore the NIRSPEC redshifts and instead use \lya\ emission and/or interstellar absorption redshifts measured from LRIS spectra for all galaxies (black squares). As discussed in Sections~\ref{Redshifts} and appendix~\ref{sec:zerrors}, the redshifts estimated from the NIRSPEC spectra have errors of $\Delta v \approx60\, \rm  km\, s^{-1}$, while the redshifts estimated from LRIS spectra typically have $\Delta v\approx130\, \rm km\, s^{-1}$. In Figure~\ref{HubbleVariations} we see that the signal appears to drop slightly more steeply for the NIRSPEC subsample, as would be expected given the smaller redshift errors, but both the NIRSPEC and pure-LRIS samples are consistent with the default sample. 

We note that for \lya\ emission and interstellar absorption we also tried using the redshift calibrations from \citet{Adelberger2005} and \citet{Steidel2010} instead of the one from \citet{Rakic2011}. For \citet{Rakic2011} the signal tends to be slightly stronger and the bootstrap errors slightly smaller, but the differences are small compared with the errors (not shown). 

One of the advantages of the POD method is the possibility to recover the optical depth in the saturated Ly$\alpha$ pixels by using higher order Lyman lines. The right panel of Figure~\ref{HubbleVariations} shows the effect of omitting this feature of the POD method. The two curves are nearly identical except for the first two bins, $0-0.18$~pMpc, where the median optical depths increase from $\sim 10$ when we make use of higher order lines to $10^4$ when we do not. The latter value is not meaningful as it is the optical depth that we assign to pixels for which saturation prevents recovery of the optical depth (see \S\ref{POD}). Without higher order lines, we cannot constrain the flux to be much smaller than the S/N ratio, which in our case corresponds to optical depths of about 4--5.

Hence, measuring the median optical depth in the circumgalactic region requires the use of higher order lines, but the recovery of the optical depth in saturated pixels appears to be unimportant at large distances. This is, however, only true if we restrict ourselves to median statistics. As we will see in Sections~\ref{cgm_scatter} and \ref{sec:coldflows}, higher order lines are also crucial at large distances if we are interested in the PDF of pixel optical depths.

\subsection{Ly$\alpha$ absorption as a function of transverse distance}

\begin{figure}
\epsscale{1.}
\plotone{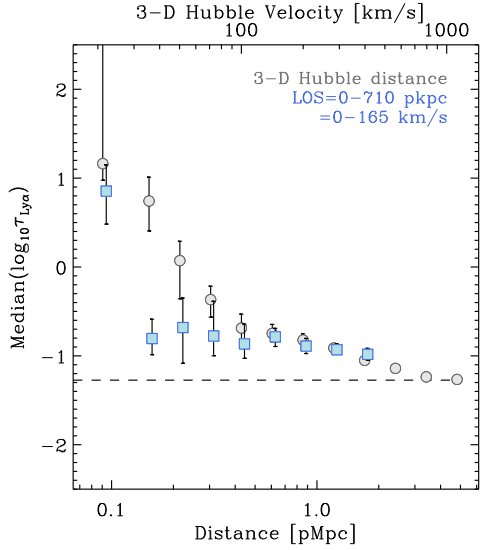}
\caption{Median (log$_{10}\tau_{\rm Ly\alpha}$) as a function of two different measures of the distance from galaxies. The blue points, which have been offset slightly for clarity, show the result as a function of transverse distance for LOS separations $< 165 ~ \rm km \, s^{-1}$ (i.e.\ 0.71~pMpc for pure Hubble flow), which is about the scale over which trends in the LOS direction are smoothed out (see Fig.~\ref{2Dcuts}) and over which errors are correlated (see Appendix~\ref{RankCorrelation}). For comparison, the grey points repeat the Hubble distance results shown in Fig.~\ref{Hubble}. The results for the true, 3-D distance are not measurable, but are expected to lie in between the transverse and Hubble results shown here. Data from this figure is tabulated in Table~\ref{tbl-3}.}
\label{Hubble_TD}
\end{figure}

Figure~\ref{2DdifferentialLOG} shows that absorption signal is smeared out over $\sim 1$~pMpc (i.e.\ $\sim 200 ~\rm km\, s^{-1}$) in the LOS direction. Indeed, for all panels of Figure~\ref{2Dcuts} the first 6 bins along the LOS (black squares), which correspond to velocities $\le 165 ~\rm km\, s^{-1}$, are consistent with each other at the $1\sigma$ level. This velocity is slightly larger than the expected redshift errors and similar to the expected circular velocities of the halos hosting our galaxies. Because there is no evidence for structure in the velocity direction on scales $\le 165 ~\rm km\, s^{-1}$ and because the errors are strongly correlated for
smaller velocity differences (see
Appendix~\ref{RankCorrelation}), it makes sense to group these first 6 LOS bins together and measure the absorption as a function of transverse distance. The result is shown as the blue squares in Figure~\ref{Hubble_TD} (all data from this figure is tabulated in Table~\ref{tbl-3}), which shows the 3-D Hubble distance results for comparison (grey circles).

Estimating the errors by bootstrapping galaxies, we find that excess absorption is detected with $\ge 2\sigma$ confidence over the full range of impact parameters. Using the more robust method of randomizing galaxy redshifts, we find that the significance is at least $2\sigma $ for all but the second point (0.13--0.18~pMpc), which is based on only 8 galaxies and for which the significance is only $1.6\sigma$. 

As expected, the Hubble and transverse results converge at large distances, where redshift space distortions are small\footnote{Note that in the last bin the absorption is slightly stronger in the transverse direction. This is a weaker version of the distortion that we attributed to the Kaiser effect when we compared the transverse and LOS directions (see Fig.~\ref{2Dcuts}).}. They also agree at very small distances (first bin; $< 0.13$~pMpc), because a small 3-D Hubble distance implies that the transverse distance must also be small (note that the reverse is not true since galaxies with small impact parameters contribute pixels with large LOS separations). On intermediate scales, however, the two measures of distance yield significantly different results, with the absorption at fixed 3-D Hubble distance being stronger than that at the same fixed transverse distance. In particular, while the transverse direction shows a rapid fall off at $b\sim 0.1~$pMpc followed by a constant excess absorption out to 2~pMpc, the absorption decreases smoothly with 3-D Hubble distance.

Because of redshift space distortions, it is not possible to measure the absorption as a function of the true, 3-D distance. However, we expect the true result to be in between the transverse and Hubble results shown in Figure~\ref{Hubble_TD}. The 3-D Hubble distance may overestimate the true distance on intermediate scales, because it assumes all velocity differences to be due to the Hubble flow, whereas in reality redshift errors and peculiar velocities will contribute. Note, however, that infall of the right magnitude could lead the 3-D Hubble distances to underestimate the true distances. On the other hand, because we group LOS separations of $0 - 165 ~\rm km \, s^{-1}$, the transverse distance will typically underestimate the true distance on intermediate scales because it implicitly assumes that the contribution of the Hubble flow is negligible up to velocity differences of $165 ~\rm km \,s^{-1}$.

However, we chose to average over this velocity difference with good reason: within $165 ~\rm km \, s^{-1}$ the signal is independent of velocity (see Fig.~\ref{2Dcuts}) and the errors are strongly correlated (Appendix~\ref{RankCorrelation}), which suggests that the velocities are not dominated by the Hubble flow. We therefore expect the transverse results to be closer to the truth, i.e. to better represent the real space absorption profile, and the excess absorption for 3-D Hubble distances $\sim 200$~pkpc to be due mostly to the inclusion of absorption around galaxies with smaller impact parameters. Indeed, we find that around these distances the scatter in the PODs is much greater if we bin in terms of 3-D Hubble distance than if we bin in terms of transverse distance (even though a velocity interval of $165 ~\rm km \, s^{-1}$ corresponds to a relatively large Hubble distance of 710~pkpc).

\begin{deluxetable}{ccc}
\tabletypesize{\scriptsize}
\tablecaption{Data from Figure~\ref{Hubble_TD}.
\label{tbl-3}}
\tablewidth{0pt}
\tablehead{\colhead{Distance\tablenotemark{a} [pMpc]} &	\colhead{Median log$_{10} \tau_{ \rm Ly\alpha}$\tablenotemark{b}}	& 	\colhead{Median log$_{10} \tau_{ \rm Ly\alpha}$\tablenotemark{c}}	}
\startdata
0.00-0.13&1.16$^{+2.84}_{-0.19}$&0.85$^{+0.30}_{-0.37}$\\
0.13-0.18&0.74$^{+0.27}_{-0.34}$&-0.81$^{+0.22}_{-0.18}$\\
0.18-0.25&0.07$^{+0.22}_{-0.43}$&-0.68$^{+0.33}_{-0.40}$\\
0.25-0.36&-0.37$^{+0.15}_{-0.20}$&-0.78$^{+0.39}_{-0.22}$\\
0.36-0.50&-0.69$^{+0.16}_{-0.18}$&-0.87$^{+0.22}_{-0.16}$\\
0.50-0.71&-0.75$^{+0.10}_{-0.09}$&-0.79$^{+0.10}_{-0.11}$\\
0.71-1.00&-0.82$^{+0.07}_{-0.07}$&-0.89$^{+0.09}_{-0.08}$\\
1.00-1.42&-0.91$^{+0.04}_{-0.04}$&-0.93$^{+0.07}_{-0.05}$\\
1.42-2.00&-1.05$^{+0.03}_{-0.03}$&-0.98$^{+0.07}_{-0.07}$\\
2.00-2.83&-1.14$^{+0.01}_{-0.03}$&\\
2.83-3.99&-1.24$^{+0.01}_{-0.03}$&\\
3.99-5.64&-1.27$^{+0.02}_{-0.01}$&\\
\enddata
\tablenotetext{a}{Distance bins in pMpc.}
\tablenotetext{b}{The median log$_{10}\tau_{\rm Ly\alpha}$ as a function of 3-D distance, with the 1$\sigma$ confidence interval.}
\tablenotetext{c}{The median log$_{10}\tau_{\rm Ly\alpha}$ a function of transverse distance for LOS separations $< 165\rm\, km\, s^{-1}$ (i.e. 0.71 pMpc for pure Hubble flow), with the 1$\sigma$ confidence interval.}
\end{deluxetable}

\subsubsection{Scatter}
\label{sec:scatter}

We have so far focused on the median absorption. However, the dashed curves in Figure~\ref{Hubble} demonstrate that there is a large degree of scatter in the distribution of PODs at a fixed distance. In this section we will therefore investigate how the median absorption around individual galaxies varies from galaxy to galaxy\footnote{We could also have studied the distribution of individual PODs, but we prefer to consider only one data point per galaxy (i.e.\ the median POD within 165~$\rm km\,s^{-1}$) because galaxies are independent, whereas pixels are not.}.

\begin{figure*}
\epsscale{1.1}
\plotone{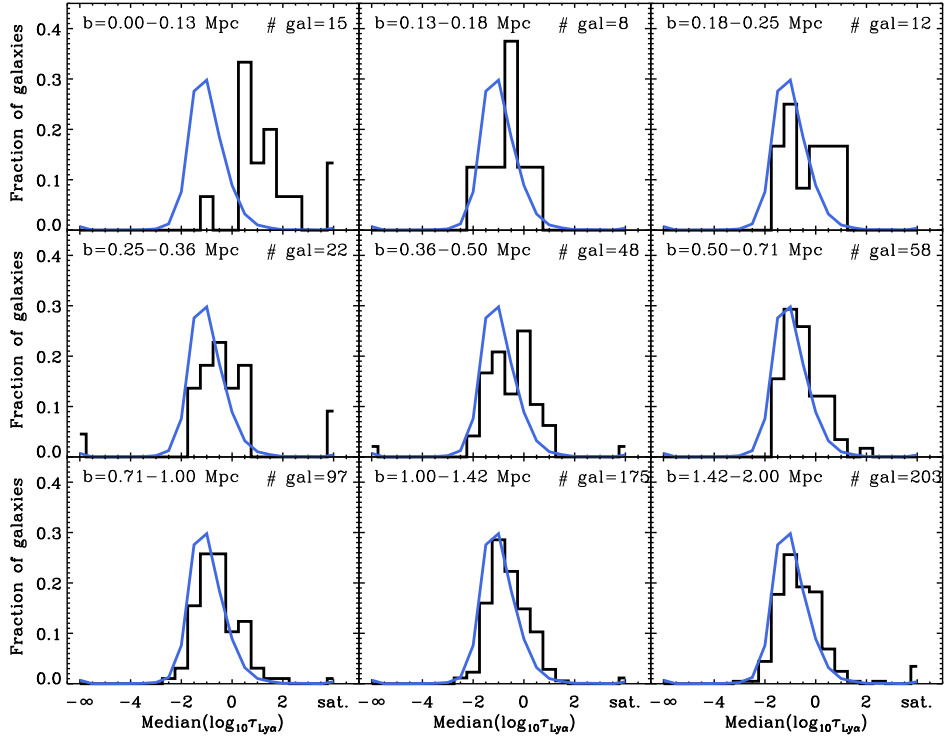}
\caption{The histograms show the distribution of the median HI \lya\ optical depth within 165~$\rm km\,s^{-1}$ from galaxies. Different panels correspond to different impact parameters in the range 0--2 pMpc, as indicated above each panel. The blue curve, which is repeated in each panel, shows the distribution after randomizing the galaxy redshifts, i.e., the distribution expected in the absence of absorption correlated with galaxy positions. There is strong ($\gtrsim 2-6\sigma$) evidence for excess absorption at all impact parameters, but there is no evidence for a trend with impact parameter beyond 0.13~pMpc.}
\label{fig:scatter}
\end{figure*}

Figure~\ref{fig:scatter} shows histograms of the distribution of the median optical depth in velocity intervals of $\pm 165~\rm km\,s^{-1}$ centered on galaxies. The different panels correspond to different impact parameters. The $x$-axis has $-\infty$ as its lower limit. This indicates pixels set to $\tau=10^{-5}$, because they had $F>1$ (see Section~\ref{POD} for details). We expect the true optical depths of these pixels to be
similar to the inverse of the S/N ratio, which is $\sim 10^{-2}$. The
label ``saturated" at the high-absorption end of the $x$-axis is for
saturated pixels whose optical depth could not be recovered using
higher order Lyman lines (because they are either also saturated or
unavailable). These pixels do not necessarily have optical depths that
are higher than the highest recovered values ($\sim 10^2$), because
higher order lines can be either contaminated or unavailable. In that
case we cannot constrain the flux to be smaller than the S/N ratio,
which corresponds to optical depths of about 4--5.

Comparing the different panels with each other, it seems that the distribution is shifted to higher optical depths for $b < 0.13$~pMpc, but that the results are similar for impact parameters $0.13 < b< 2.0$~pMpc. Indeed, Kolmogorov-Smirnov (K-S) tests show that, except for the first panel, panels 2--8 are consistent with the last panel (at the $1\sigma$ level). For $b< 0.13$~pMpc, however, the distribution of median optical depths differs at the 99.98\% confidence level ($3.7\sigma$). These results are consistent with our findings for the medians of the distributions (Fig.~\ref{Hubble_TD}, blue squares).

Ignoring the outliers for which the median optical depth is saturated or $-\infty$, the scatter is about 0.75~dex. The median absorption near galaxies with a fixed impact parameter is thus highly variable, suggesting that the gas is clumpy. This is true at all impact parameters and therefore not a distinguishing feature of the circumgalactic medium. 

The blue curve, which is repeated in every panel, indicates the distribution after randomizing the galaxy redshifts. It therefore gives the distribution that we would expect to measure if the absorption were uncorrelated with galaxy positions. Using the K-S test to compare each histogram with the blue curve, we find that except for the second panel (which contains only 8 galaxies), all are discrepant at the 2$\sigma$ to 6$\sigma$ level. For $b>0.5$~pMpc the excess absorption appears small, but the difference is in each case significant at the $\gtrsim 4\sigma$ level. For the second panel ($b=0.13-0.18$~pMpc) the significance of the detection of excess absorption is $\lesssim1\sigma$. These results are in near perfect agreement with those based on median statistics. Thus, examination of the full distribution of median optical depths around galaxies confirms the result obtained for the medians. There is a sharp drop in the absorption around $10^2$~pkpc, which is similar to the virial radius, followed by a near constant, small, but highly significant, excess absorption out to transverse separations of at least 2~pMpc.

\subsection{Interpreting PODs}\label{Interpretation}

\begin{figure*}
\resizebox{0.33\textwidth}{!}{\includegraphics{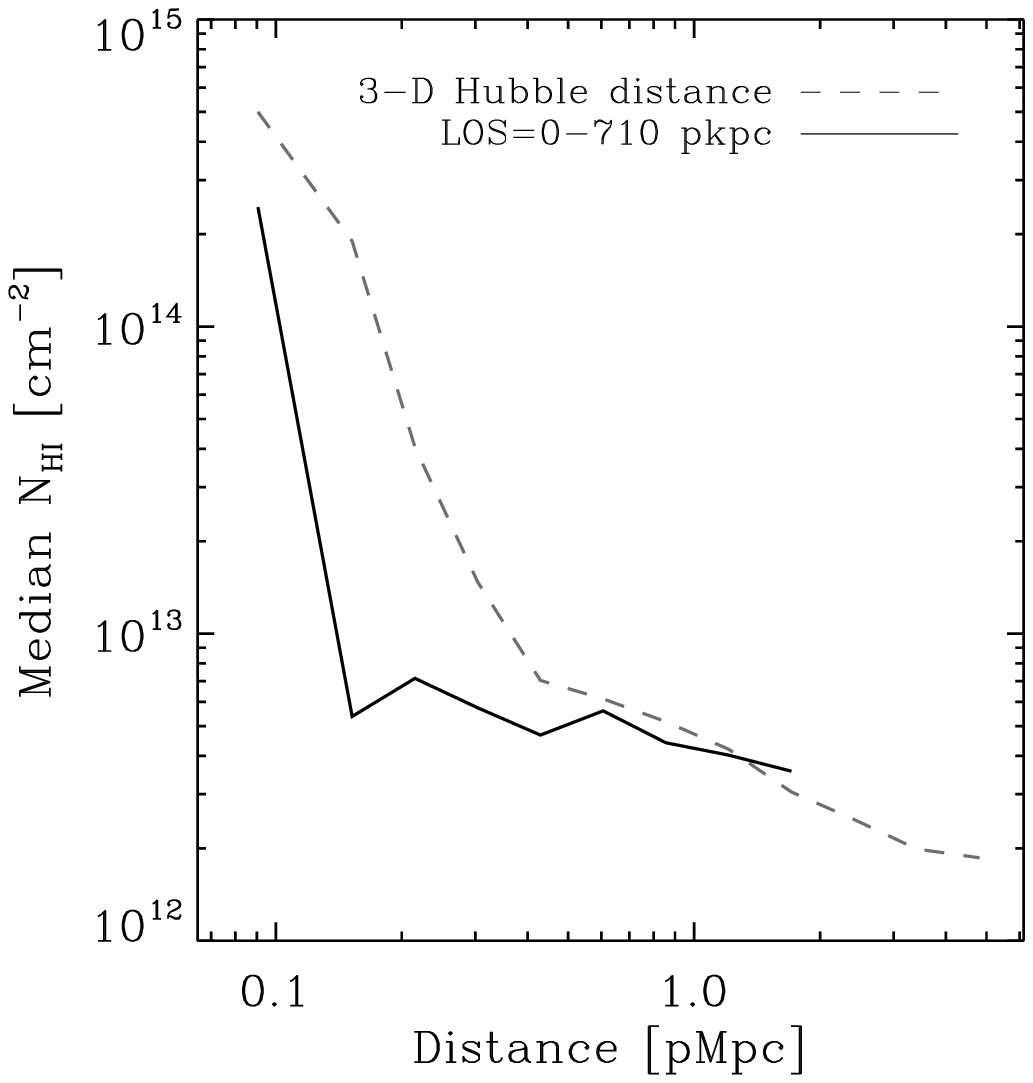}}
\resizebox{0.33\textwidth}{!}{\includegraphics{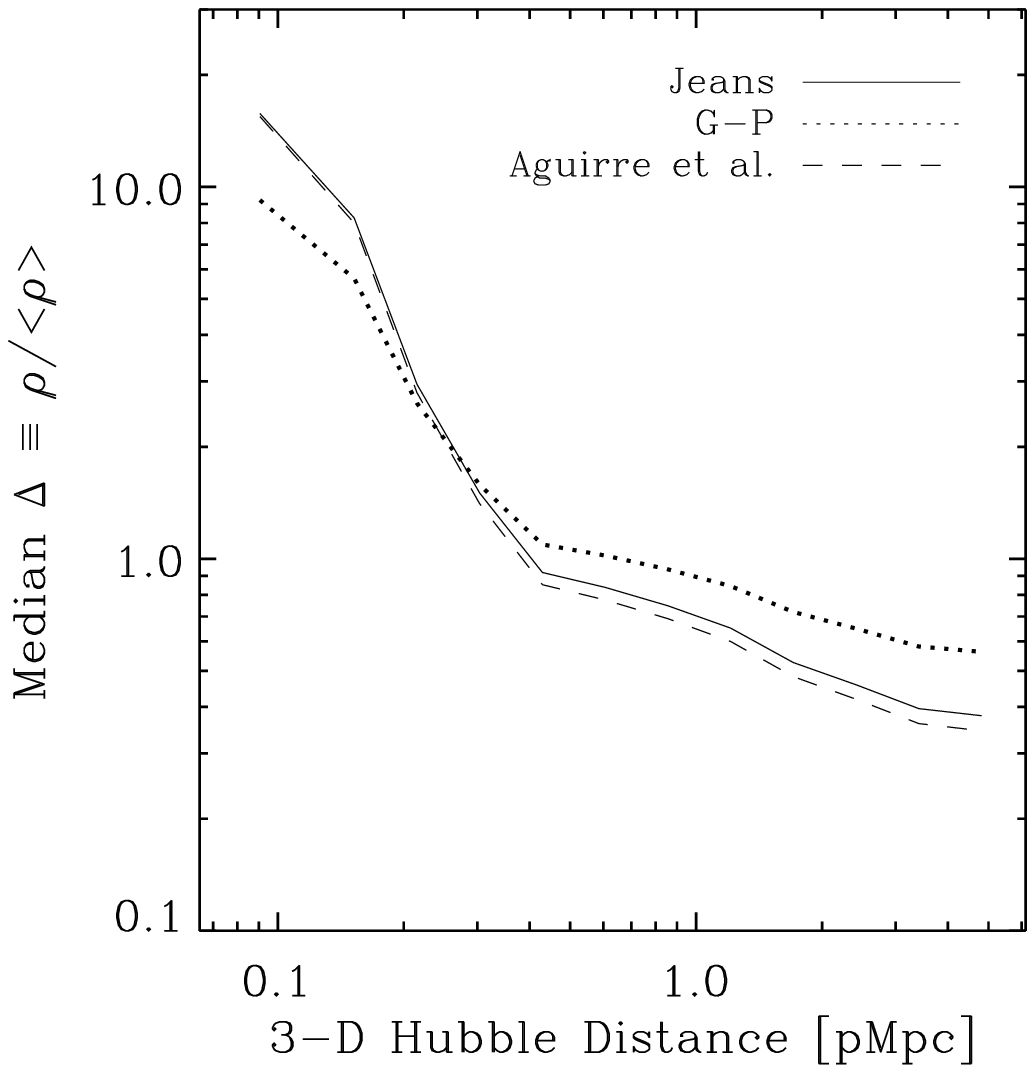}}
\resizebox{0.33\textwidth}{!}{\includegraphics{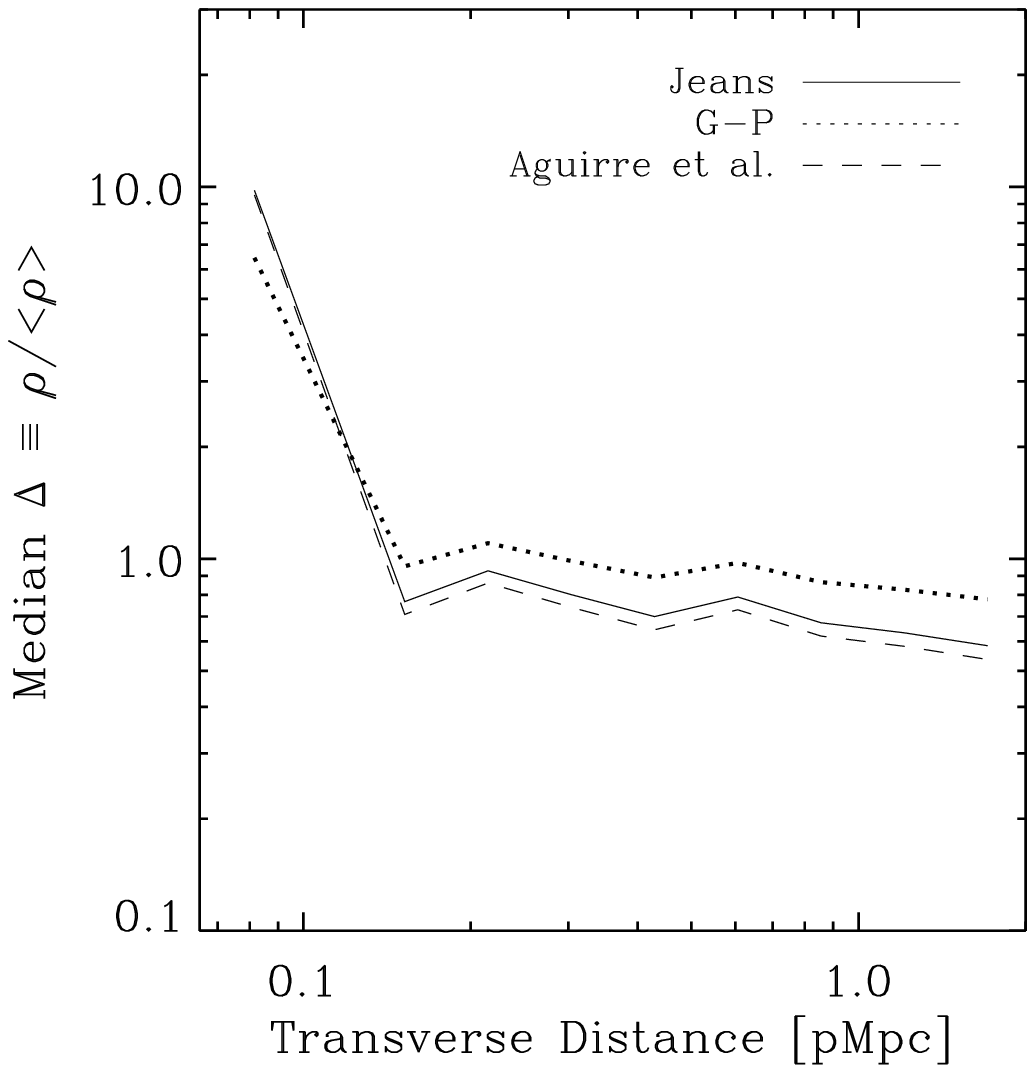}}
\caption{Inferred median neutral hydrogen column density (left panel) and inferred median overdensity (middle and right panels) as a function of distance from galaxies. Column densities were obtained from the measured, median \lya\ optical depths and equation~(\ref{EquationTau}). Overdensities were obtained by converting the median optical depths using the \citet{Schaye2001} ``Jeans'' approximation (solid curves), the fluctuating Gunn-Peterson approximation (dotted lines), or the fit to the results of the hydrodynamic simulation of \citet{Aguirre2002} (dashed curves). Overdensities inferred from observed optical depths are effectively volume weighted (and are lower than mass-weighted overdensities). The right panel and the solid curve in the left panel show results as a function of transverse distance for velocity separations $< 165 ~\rm km \, s^{-1}$, while the middle panel and the dashed curve in the left panel are for 3-D Hubble distances. Typical column densities (overdensities) decrease from 
$\sim 10^{14.5} ~ \rm cm^{-2}$ ($\sim 10$) at $\sim 10^{-1}~ \rm pMpc$ to  $\lesssim 10^{13} ~ \rm cm^{-2}$ ($\lesssim 1$) at  $\gtrsim 1 ~\rm pMpc$. }\label{HubbleOverdensity}
\end{figure*}

The central optical depth is related to the absorbing gas column density, $N$, through the following approximate relation:
\[
\tau_{\rm 0}\approx \left(\frac{N}{3.43\times 10^{13} \rm \, cm^{-2}}\right)\left(\frac{\it f}{0.4164}\right)
\]
\begin{equation}
\times \left(\frac{\lambda_{\rm 0}}{1215.67\,\rm \AA}\right)\left(\frac{b_{\rm D}}{26\, \rm km\, s^{-1}}\right)^{-1}	
\label{EquationTau}
\end{equation}
where $f$ is the oscillator strength, $\lambda_{\rm 0}$ is the transition's rest wavelength, $b_{\rm D}=\sqrt{2}v_{\rm RMS}=(\rm FWHM)/2\sqrt{\rm ln2}$ is the line width, and $v_{\rm RMS}$ is the line of sight velocity dispersion \citep[e.g.][\S9.5]{Padmanabhan2002}. In our study we use statistics based on either median optical depths or (in \S\ref{sec:coldflows}) on the maximum optical depth within a given distance range from a galaxy.  Measurements based on the maximum optical depth in a given region are relatively easy to interpret using Equation~(\ref{EquationTau}). However, given that most of our analysis is based on median optical depths, all the conversions that we make to column densities of the absorbing gas are likely to be underestimates if most of the lines are not blended, and could be overestimates in regions where the lines are highly blended. For example, if there is a single line in a given velocity interval, the column density estimated from the median optical depth would be lower than the real column density of such a system. If, on the other hand,  there is one strong line and a number of weaker lines that are blended with it, then the column density inferred from the median optical depth could be higher than the median column density of individual absorption systems.

The left panel of Figure~\ref{HubbleOverdensity} shows the median neutral hydrogen column density as a function of both transverse distance for LOS velocities $0-165 ~\rm km \, s^{-1}$ (solid curve) and 3-D Hubble distance (dashed curve), which we obtained from Figure~\ref{Hubble_TD} and equation~(\ref{EquationTau}), using the typical line width of $26\rm\, km\, s^{-1}$  measured for our sample by \citet{Rudie2012}. The median column density decreases from $\sim 10^{14.5}\,{\rm cm}^{-2}$ at $\sim 10^2$~pkpc to $\sim 10^{12.7}\,{\rm cm}^{-2}$ at $\sim 1$~pMpc, which is in excellent agreement with results from \citet{Rudie2012} based on Voigt profile decompositions. Note that if we had selected the strongest system within a given distance of each galaxy, instead of taking into account all pixels within that distance, the
 value of the median column density (i.e., the median of the maximum column density) would have been
 significantly higher.

To gain intuition about what the observed absorption represents, we will convert the optical depths into overdensities using two approximations. Combining Equation~(\ref{EquationTau}) with Equation~(10) of \citet{Schaye2001}, who treats Ly$\alpha$ absorbers as gravitationally confined gas clouds with sizes of order the local Jeans length, we obtain:
\begin{eqnarray} 
\Delta&\approx &2.1 \tau_{\rm 0,Ly\alpha}^{2/3} \Gamma_{12}^{2/3} \left(\frac{1+z}{3.36}\right)^{-3} \left (\frac{T}{2\times 10^4\,{\rm
K}}\right )^{0.17} \nonumber \\  
&& \times \left(\frac{f_{\rm g}}{0.162}\right)^{-1/3}\left(\frac{b}{26\, \rm km\, s^{-1}}\right)^{2/3}
\label{EquationJeans}
\end{eqnarray}
where $\Delta \equiv \rho / \langle\rho \rangle$ is the density of gas in units of the mean baryon density of the Universe, $\Gamma_{12}$ is the photo-ionization rate in units of $10^{12}\,\rm s^{-1}$, and $f_{\rm g}$ denotes the fraction of the cloud mass in gas.
We have assumed a temperature typical for the moderately overdense IGM \citep[e.g.][]{Schaye2000a,Lidz2010,Becker2011}, a line width consistent with the median value measured by \citet{Rudie2012} for our data, and a photo-ionization rate appropriate for ionization by the ultraviolet background radiation \citep[e.g.][]{Bolton2005,FG2008}. In collapsed gas clouds $f_{\rm g}$ could be close to unity, but far away from galaxies gas will not be in dense clumps and $f_{\rm g}$ should be close to its universal value of $\Omega_{\rm b}/\Omega_{\rm m}$. Note that these densities are effectively evaluated on the local Jeans scale, which is typically $\sim10^2\,\rm pkpc$ for the densities of interest here \citep[][]{Schaye2001}.

The above expression is a good approximation for overdense absorbers. Absorbers with densities around or below the cosmic mean have not had sufficient time to reach local hydrostatic equilibrium \citep{Schaye2001} and will be better described by the fluctuating Gunn-Peterson approximation \citep[e.g.][]{Rauch1997}, which assumes smoothly varying density fluctuations and pure Hubble flow, yielding
\begin{equation} 
\Delta\approx 2.02 \tau_{\rm Ly\alpha}^{1/2} \Gamma_{12}^{1/2} \left(\frac{1+z}{3.36}\right)^{-9/4} \left (\frac{T}{2\times 10^4\,{\rm K}}\right )^{0.38}.
\label{EquationGP}
\end{equation}

Both equations (\ref{EquationJeans}) and (\ref{EquationGP}) assume primordial abundances, highly ionized gas, and photo-ionization equilibrium. However, close to galaxies, UV radiation from local sources may dominate over the background and the gas may be shock-heated to temperatures sufficiently high for collisional ionization to dominate. Both of these effects would cause us to underestimate the gas density, possibly by a large factor.

The middle and right panels of Figure~\ref{HubbleOverdensity} show the median overdensity profiles obtained after applying the above equations to the median optical depth measured as a function of 3-D Hubble and transverse distance (for LOS separations $0-165 ~\rm km \, s^{-1}$), respectively. The ``Jeans'' and fluctuating Gunn-Peterson approximation are generally in good agreement, although the former yields steeper density profiles at small distances, where the gas is highly overdense. However, in this regime we do not expect the fluctuating Gunn-Peterson approximation to hold. The dashed curve shows the result if we convert the recovered optical depths into overdensities using the fit to hydrodynamical simulations given in \citet{Aguirre2002} for $z = 2.5$. The relation provided by those authors was obtained by producing mock Ly$\alpha$ absorption spectra for sight lines through a hydrodynamical simulation, scaling the simulated spectra to fit the observed mean flux decrement, and then fitting the relation between the Ly$\alpha$ optical depth-weighted overdensity of gas responsible for the absorption in each pixel and the recovered Ly$\alpha$ POD in that pixel. The agreement between the hydrodynamical simulation and the Jeans method is clearly excellent.

We conclude that typical gas overdensities decrease from $\gtrsim 10$ at distances $ \lesssim 10^2$~pkpc, which is similar to the virial radii of the halos hosting our galaxies, to $\sim 1$ at $\sim 1$~pMpc. The steepness of the drop at intermediate scales is uncertain due to redshift space distortions, but it is likely to be bracketed by the the middle and right panels of Figure~\ref{HubbleOverdensity}. Observe that on large scales the overdensity asymptotes to values smaller than unity. This is expected, as the median optical depth is a volume weighted measure rather than a mass-weighted measure. While the mass-weighted mean overdensity is by definition unity, the volume-weighted mean overdensity is smaller because underdense regions dominate the volume. As most of the pathlength in the Ly$\alpha$ forest is across voids, the median density inferred from the Ly$\alpha$ forest is less than 1. The difference between volume and mass-weighted quantities probably also accounts for the relatively low overdensity measured around the virial radius ($\sim 10^{-1}$~pMpc).

\subsection{Circumgalactic matter}\label{CGM}

The smallest impact parameter in our sample is $\approx55$ pkpc. It is difficult to probe scales smaller than  this with QSO-galaxy pairs owing to the small number of bright QSOs and the rarity of close pairs, as well as to the difficulty of observing objects right in front of a bright QSO. For such small scales it is therefore more efficient to resort to using galaxies as background objects \citep[e.g.][]{Adelberger2005,Rubin2009,Steidel2010}. Nevertheless, in this section we will use our sample of background QSOs to study the gas within transverse distances of 200 pkpc of galaxies. But before doing so, we will compare our results with the small-scale data from \citet{Steidel2010}.

\subsubsection{Comparison with results for galaxy-galaxy pairs}
 
 \begin{figure}
\resizebox{0.5\textwidth}{!}{\includegraphics{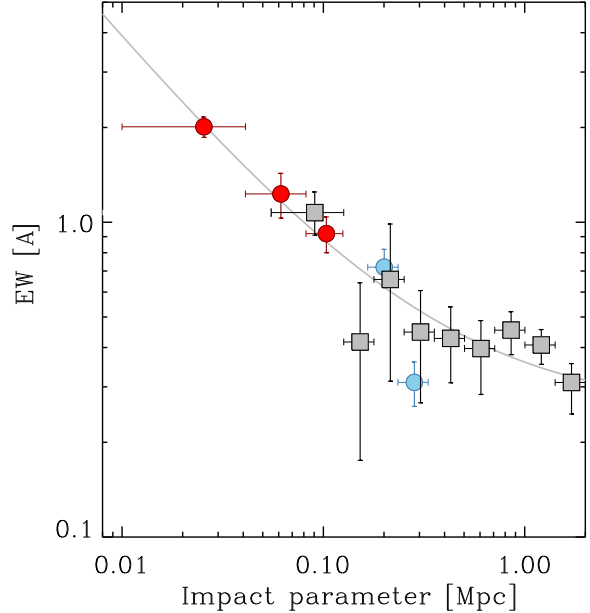}}
\caption{Rest-frame equivalent width of Ly$\alpha$ absorption as a function of impact parameter, for the galaxy-galaxy pair samples from \citet{Steidel2010} shown as red circles, and as measured from the data set used in this paper, QSO-galaxy pairs, as grey squares. The blue circles show QSO-galaxy pair results from \citet{Steidel2010}, using the same KBSS data, but a different way of measuring EW and different bin sizes.) The vertical error bars are estimated by bootstrap resampling the galaxy sample. The horizontal error bars indicate bin sizes. The grey line shows the best fit: EW$=0.11b^{-0.76}+0.25$, where $b$ is impact parameter of galaxies. }\label{EW}
\end{figure}

 \citet{Steidel2010} measured Ly$\alpha$ rest-frame equivalent width (EW) as a function of distance from the same sample of galaxies that we study here. Since they used background galaxy spectra for probing foreground galaxies' circumgalactic matter, they were able to study smaller  scales than we can probe with the QSO spectra. While they could measure the EW in a stack of background galaxy spectra, the spectral resolution of the galaxy spectra was too low to measure column densities or resolved pixel optical depths and the S/N was too low to obtain EW measurements for individual galaxies. The red circles in Figure~\ref{EW} show their measurements at small impact parameters and the blue circles show their measurements using QSO spectra, together with our EW measurements from QSO spectra at larger impact parameters (black squares). Although we used the same data, their technique for measuring EW differs from ours \citep[see][for more details]{Steidel2010}.
 
We computed the EWs as follows. We shifted the spectra into the rest-frame of each galaxy within a given impact parameter bin, found the mean flux profile, divided it by the mean flux level of all pixels in the spectra (i.e.\ 0.804) in order to mimic the effect of continuum fitting low-quality spectra, and integrated the flux decrement over $\pm500\, \rm km\, s^{-1}$ centered on the galaxies' positions. We verified that using a larger velocity interval ($\le1000 \,\rm km\, s^{-1}$) gives consistent results.
 
It appears that the absorption strength falls off according to a power law, out to $\sim800$ pkpc. Beyond this impact parameter the relation flattens off at EW$\sim0.2-0.3$~\AA.

\subsubsection{Scatter}
\label{cgm_scatter}

\citet{Adelberger2005} found that the absorption near galaxies follows a bimodal distribution. While the flux decrement is usually large, $\approx1/3$ of galaxies are not associated with strong Ly$\alpha$ absorption within a 3-D Hubble distance of 1~$h^{-1}$~pMpc (i.e.\ $\approx400\, \rm pkpc$). 

As shown in Figure~\ref{fig:scatter}, we also find that the absorption varies strongly from galaxy to galaxy. However, we showed that this is not only true near galaxies, but also if we look in random places of the spectrum. Moreover, the distribution of median optical depths is not bimodal. 
\begin{figure*}
\resizebox{0.45\textwidth}{!}{\includegraphics{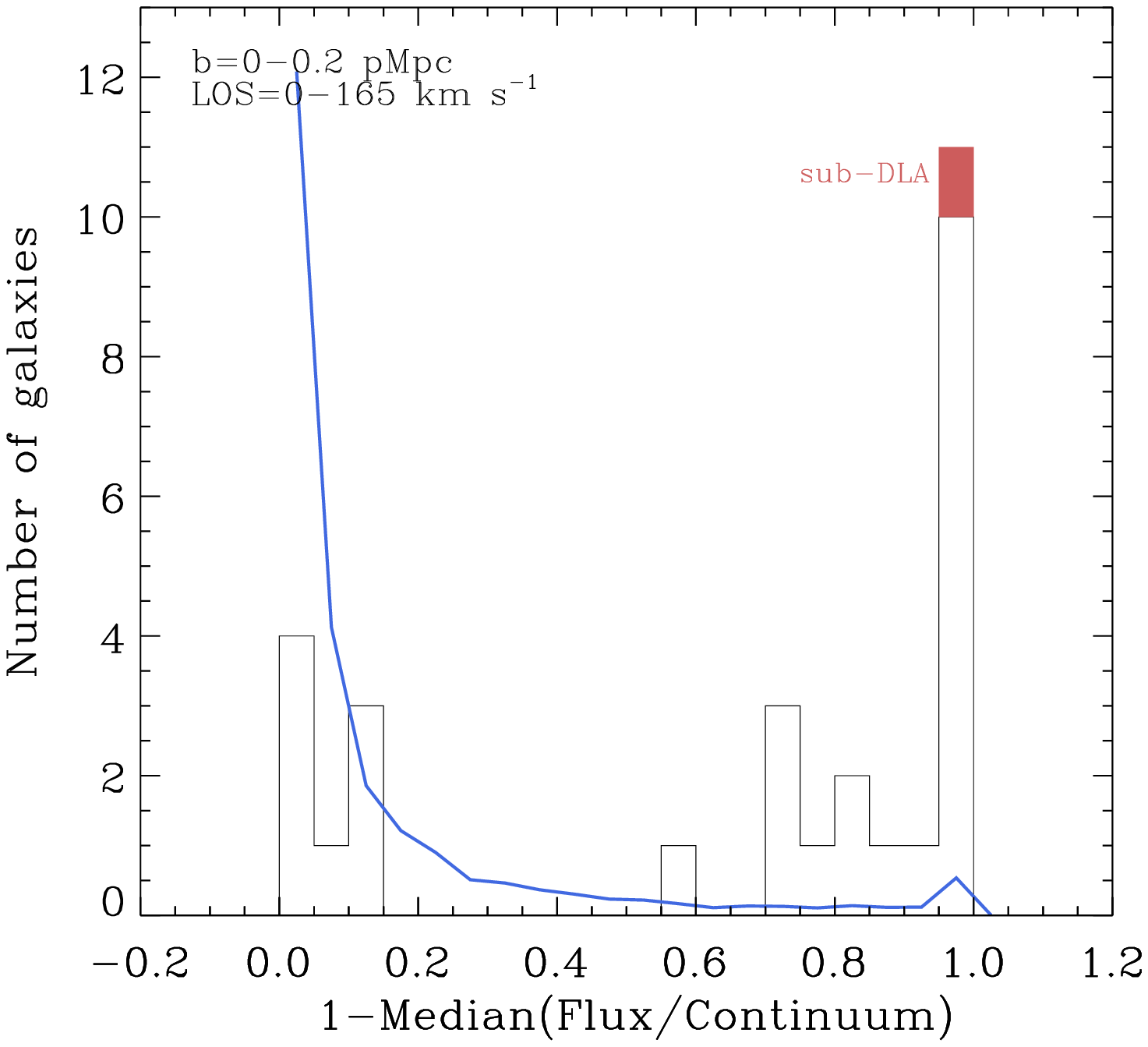}}
\resizebox{0.45\textwidth}{!}{\includegraphics{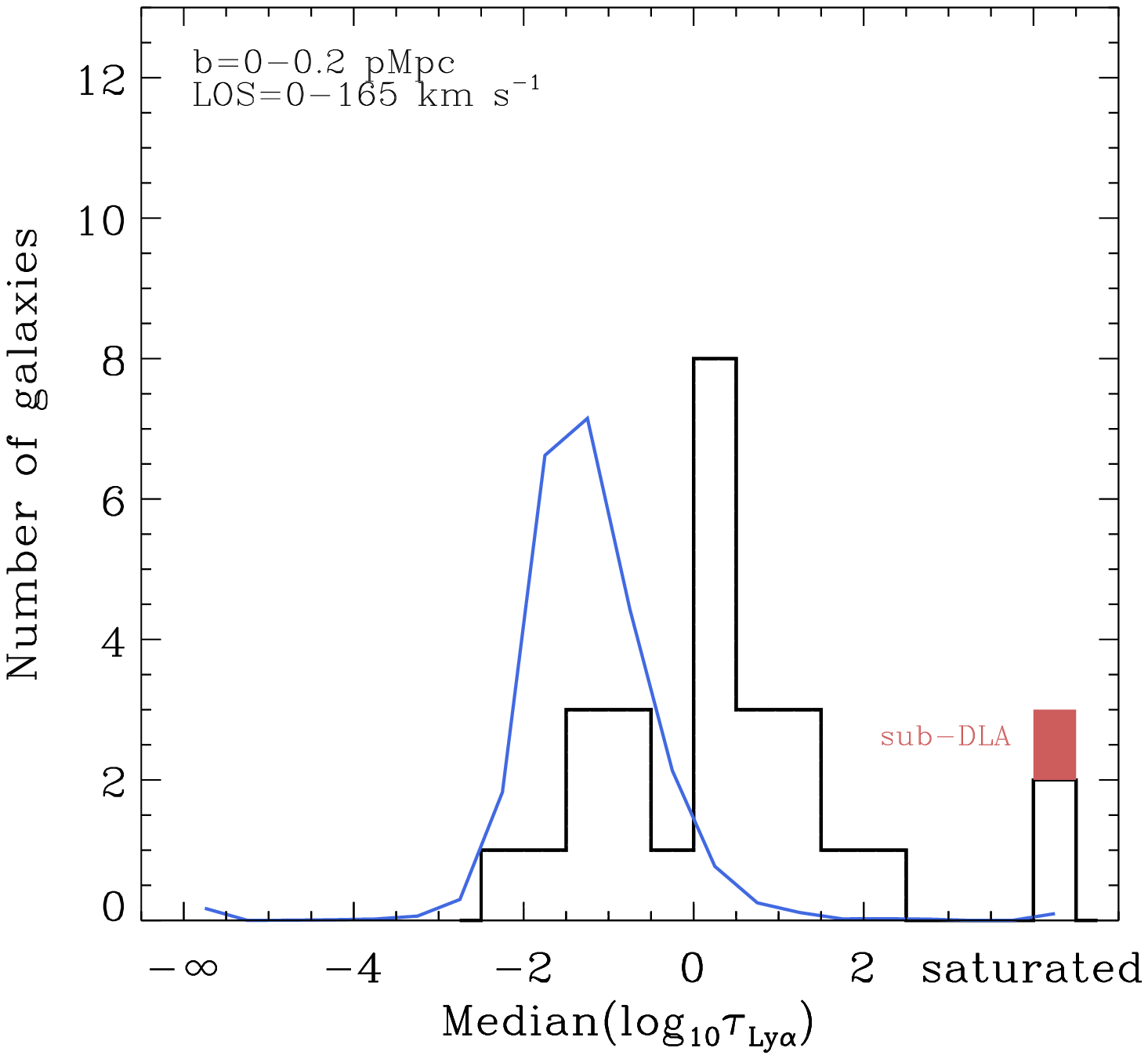}}
\caption{Number of galaxies vs.\ median flux (left panel) and median log$_{10}\tau_{\rm Ly\alpha}$ (right panel) within 165 $\rm km\, s^{-1}$ from galaxies with impact parameters $b<200\, \rm pkpc$ from galaxies. The blue curves show the results expected if absorbers and galaxy positions were uncorrelated, and were obtained after randomising the galaxy redshifts 1,000 times. In the right panel $-\infty$ indicates the pixels with $F>1$, and ``saturated" indicates saturated pixels for which the optical depth could not be recovered  (see text for more details). The observed absorption is enhanced at the $>6\sigma$ level.}\label{HistogramGalaxy200kpc}
\end{figure*}

As \citet{Adelberger2005} did not measure optical depths, we show the distribution of the median flux decrement for impact parameters $< 200\, \rm pkpc$
and velocity differences $< 165\, \rm km\, s^{-1}$ from galaxies in the left panel of Figure~\ref{HistogramGalaxy200kpc}. For comparison, the right panel shows the corresponding median optical depths. As before, we chose this velocity interval because there is little structure in the median profiles for velocity separations smaller than this value (see Figs.~\ref{2DdifferentialLOG} and \ref{2Dcuts}), which is not surprising given that $165\, \rm km\, s^{-1}$ is similar to both the redshift errors and the circular velocities of the halos thought to host the galaxies.  The blue curves show the results for randomized galaxy redshifts, where we assign random redshifts (within the Ly$\alpha$ forest redshift range) to the galaxies in our sample. The curves are the result of estimating a histogram for randomized redshifts 1,000 times, and taking the mean median in each bin. A K-S test shows that the two distributions are discrepant at the $>6\sigma$ level, so the absorption is clearly enhanced near galaxies. 
  
It can be seen in the left panel that galaxies either show very little
transmission, or relatively high flux. The flux distribution is thus bimodal, in agreement with \citet{Adelberger2005}. However, the right panel shows that there is no
evidence for bimodality of the optical depth distribution. Hence, the bimodality seen in the left panel is a consequence of the mapping $F=\exp(-\tau)$ which bunches the low (high) optical depth tail of the distribution together at a flux decrement of zero (one). Note that the shape of the optical depth distribution (which appears to be approximately lognormal) is physically more relevant than that of the flux, because the former is proportional to the neutral hydrogen column density.

We will demonstrate next that the scatter is not random as the optical depth is strongly anti-correlated with the impact parameter.

\subsubsection{Cold flows}
\label{sec:coldflows}

\begin{figure*}
\resizebox{0.33\textwidth}{!}{\includegraphics{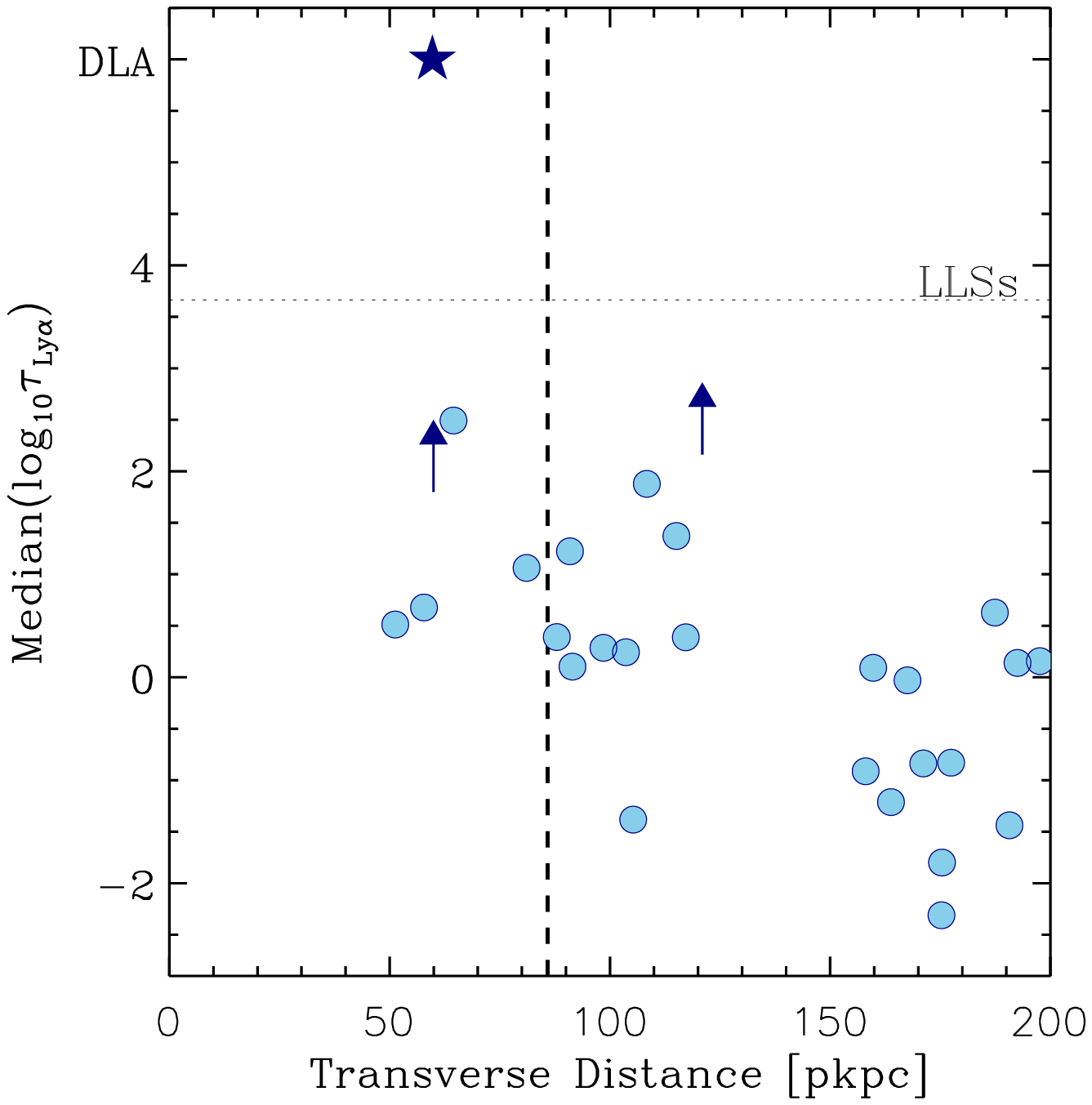}}
\resizebox{0.33\textwidth}{!}{\includegraphics{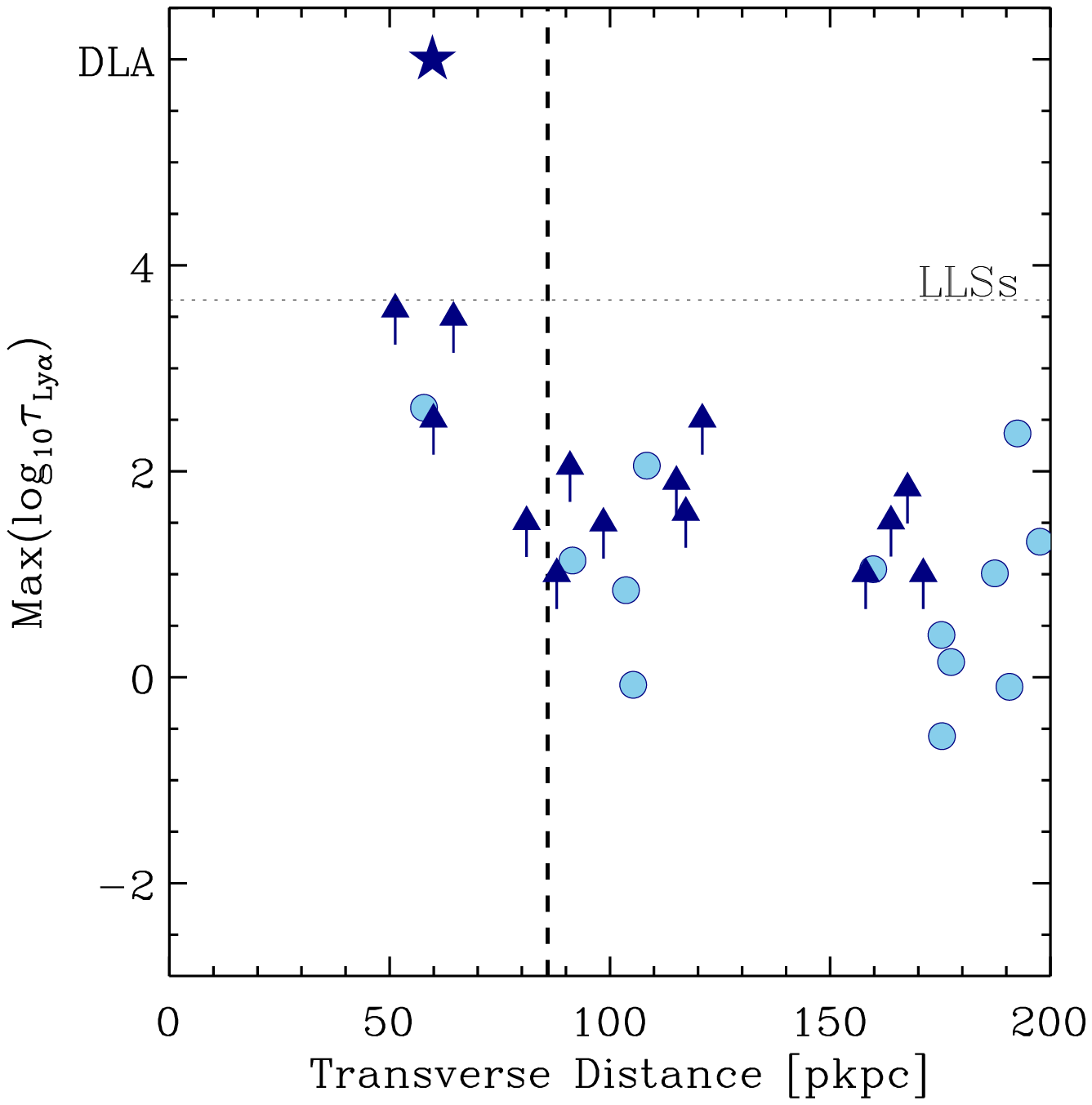}}
\resizebox{0.33\textwidth}{!}{\includegraphics{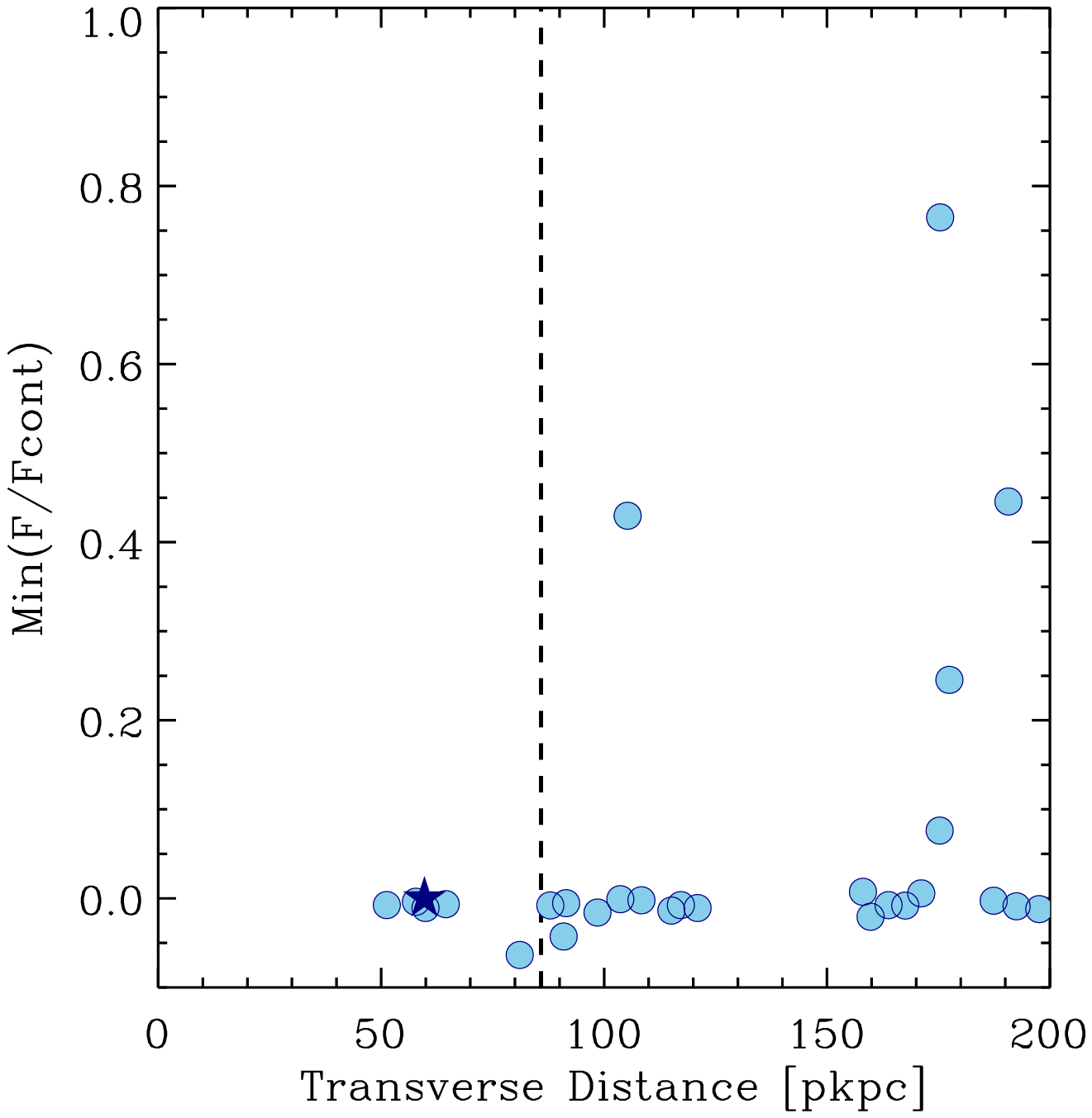}}
\caption{\emph{The left and middle panels} show, respectively, the median and the maximum optical depth within 165 $\rm km\, s^{-1}$ from each galaxy that is separated from the LOS by less than 200~pkpc as a function of its impact parameter. In cases where the median/maximum corresponded with a saturated pixel without an optical depth estimate we plot the largest recovered value of the POD within 165 $\rm km\, s^{-1}$ as an upward pointing arrow - these are lower limits to the actual optical depths.  The dotted horizontal line indicates the approximate threshold for Lyman limit absorbers. Stars indicate a sub-DLA, identified via its damping wings (and with line centers $< 165 ~\rm ~ km\, s^{-1}$ from the  nearest galaxy). \emph{The right panel} shows the minimum flux within 165 $\rm km\, s^{-1}$ from each galaxy.  The vertical dashed lines indicate the virial radius for typical halos hosting the galaxies in our sample, with $M_{\rm h}=10^{11.9}\rm M_{\odot}$ at $z=2.4$.}\label{AbsorptionProfile250kms}
\end{figure*}

In recent years cosmological simulations and theoretical models have converged on the idea that gas accretion is bimodal. While the gas accreting onto massive halos is shock-heated to the virial temperature, in lower mass halos most of the gas falls in cold ($T\sim 10^4~$K). The cold mode feeds the galaxy through filaments (cold streams) whereas the hot mode results in the formation of a hydrostatic halo which fuels the galaxy through a cooling flow \citep[e.g.][]{Keres2005,Ocvirk2008,Dekel2009,Crain2010,Voort2011a,Voort2011b,FG2011}. At $z=2$ halo masses of $\sim 10^{12}~\rm M_\odot$, i.e.\ similar to those hosting our galaxies (Trainor et al. 2011, in preparation), are especially interesting in that they mark the transition between the cold- and hot-mode accretion regimes. For example, \citet{Voort2011a} predict that the hot mode contributes on average $\approx 50\%$ and $\approx 30\%$ to the growth of $10^{12}~\rm M_\odot$ halos and their central galaxies, respectively. The simulations predict that individual galaxies in such halos are fed simultaneously through both modes with cold streams penetrating hot, hydrostatic halos. 

Despite the theoretical consensus, there is no direct observational evidence for this galaxy formation picture as of yet. Absorption by neutral hydrogen is a promising way to observe the cold streams.
Based on a high-resolution simulation of a single halo with mass $3 \times 10^{11}~\rm M_\odot$ at $z=2$, \citet{FG2010} predict that the covering fraction of cold flows with Lyman Limit (LLS; $10^{17.2}\leq N_{\rm HI}\leq2\times10^{20}\, \rm cm^{-2}$) and DLA column densities ($N_{\rm HI}>2\times10^{20}\, \rm cm^{-2}$), within the virial radius (2 virial radii) is 10--15\% (4\%) and 3--4\% (1--2\%), respectively, where they quote 74~pkpc for the virial radius. \citet{Kimm2011} use cosmological simulations to predict that the covering fraction of systems with $N_{\rm HI}>10^{20}\, \rm cm^{-2}$ within 100~pkpc of halos of $\sim 10^{12}~\rm M_{\odot}$ at $z=2.5$ is $\approx5$\%. \citet{Stewart2011} simulate two halos with masses of $\sim 10^{11.5}~\rm M_\odot$ at $z=2$ and predict covering fractions for $N_{\rm HI}>10^{16}\, \rm cm^{-2}$ of about 15\% at 33~pkpc. \citet{Fumagalli2011} study the covering fraction of cold streams around 7 haloes. For two haloes with virial masses of $5.9\times10^{11}\rm M_{\odot}$ and $9.2\times10^{11}\rm M_{\odot}$ at $z=2.33$, which are typical for star-forming galaxies from our sample, they predict a covering fraction for DLAs (LLSs) within 2R$_{\rm vir}$ of 2.22 (6.89) and 1.54 (9.11) per cent, respectively.

The left panel of Figure~\ref{AbsorptionProfile250kms} shows the median optical depth within $\pm 165 ~\rm km\, s^{-1}$ from galaxies, as a function of the impact parameter. The results are insensitive to the exact velocity interval chosen, although a galaxy with $b=155$ pkpc would have been a DLA system if we had used a maximum velocity separation $\ge 200 ~\rm km\,s^{-1}$. Each circle represents a galaxy. Sub-DLA systems are shown as star symbols.
In two non-DLA cases the median optical depth could not be measured because more than half of the pixels were saturated and there were insufficient higher order lines to recover the optical depth. For those cases we plot the largest recovered optical depth and show it as an upwards-pointing arrow to indicate that it is a lower limit. Note, however, that there is no evidence for DLA absorption associated with any of the galaxies plotted as lower limits.
The vertical dashed line shows the virial radius for halos with mass
$\rm M=10^{11.9}\rm M_{\odot}$ (Trainor et al. 2012). To compare with predictions from the literature, it is more appropriate to look at the maximum pixel optical depth in the $\pm 165 ~\rm km\, s^{-1}$ interval around each galaxy, because we can convert this into a neutral hydrogen column density using equation (\ref{EquationTau}). This results in significantly more lower limits, as can be seen in the middle panel of Figure~\ref{AbsorptionProfile250kms}. Finally, the right panel shows the minimum flux. 

Figure~\ref{AbsorptionProfile250kms} shows a clear trend of increasing absorption strength with decreasing distance. While the scatter in the optical depth is large at all impact parameters, in terms of flux there is actually little variation within 100~pkpc. In particular, it is striking that all 10 galaxies with impact parameters smaller than 100~pkpc are associated with saturated absorbers ($F\approx0$). 

Within 100 pkpc 10 out of 10 galaxies have median $\tau_{\rm Ly\alpha} > 1$, which corresponds to a covering fraction of $100^{+0}_{-32}\%$. Within 200 pkpc this decreases to $66\pm 16\%$. If we use the maximum optical depth then the covering fraction of $\tau_{\rm Ly\alpha} > 1$ absorbers within 200 pkpc increases to $89^{+11}_{-18}\%$. Note that the central optical depth is unity for lines with column density $N_{\rm HI} \approx 4\times 10^{13}~\rm cm^{-2}$ (see equation \ref{EquationTau}).

It appears that 1--5 out of 6 galaxies  with impact parameters smaller than the virial radius of their host halos are associated with LLSs (the horizontal dashed line shows the central optical depth for a $N_{\rm HI} = 10^{17.2}\, \rm cm^{-2}$ absorber assuming a line width of $26~\rm km\, s^{-1}$), i.e. there is one secure LLS and 4 possible LLSs given that we have only lower limits on the optical depth of these absorbers. In addition, 1 out of 6 galaxies has a sub-DLA within $R_{\rm vir}$. Within two virial radii 1--15 out of 21 galaxies are associated with LL systems and 1 out of 21 galaxies is associated with a sub-DLA (in addition there would have been 1 DLA if we had used a maximum velocity separation $\ge 200 ~\rm km\,s^{-1}$).
These numbers are higher than the predictions from \citet{FG2010} and \citet{Kimm2011}, and also higher than \citet{Fumagalli2011} for DLAs (they are consistent with their results for LLSs),  but the errors are too large for the discrepancy to be significant. We note that when it comes to measuring the column densities of strongly saturated systems (e.g.\ Figure~\ref{AbsorptionProfile250kms} where our method yields a number of lower limits), it would be more appropriate to use Voigt profile decompositions, as are presented by \citet{Rudie2012}. This other method gives results fully consistent with those presented here.

Even though these observations are consistent with what is seen in simulations, it is unclear whether the observed absorbing gas is inflowing or outflowing. \citet{Steidel2010} found, using 500 galaxy pairs sampling angular scales of 1--15'', that the circumgalactic medium within $\approx80-90$ pkpc shows strong \ion{H}{1} and low-ion metallic absorption, which is consistent with the findings in this paper.  However, they also find that galaxy spectra exhibit kinematics consistent with radial flows with velocity increasing outward.

\section{The distribution of galaxies around absorbers}\label{IGMCentricView}
In the previous sections we investigated the typical IGM environment of the galaxies in our sample. Here we address the complementary question: what is the galaxy environment for a pixel of a given optical depth? 

\begin{figure}
\includegraphics[height=3in,trim=1.5cm 0.5cm 1.5cm 1.5cm,clip]{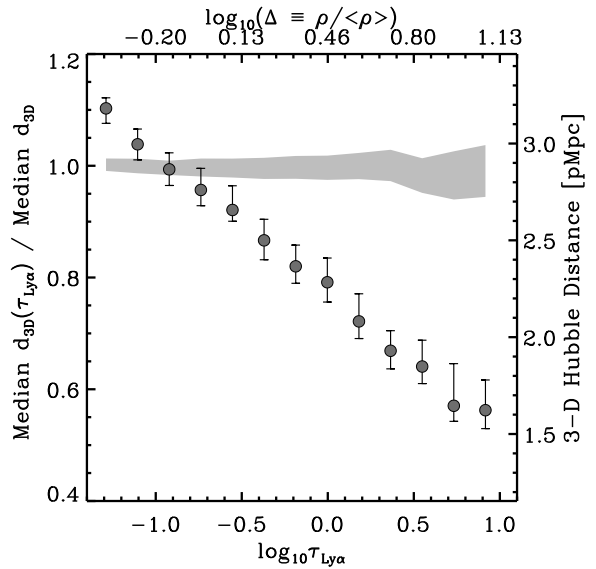}
\caption{The median 3-D Hubble distance to the nearest galaxy in our sample, normalized by the median 3-D Hubble distance to the nearest galaxy for all pixels, as a function of the pixel's Ly$\alpha$ optical depth. We normalize by this  distance in order to get results that do not depend on the galaxy sample's completeness. The unnormalized distance scale is shown on the right $y$-axis. The grey shaded area shows the $1\sigma$  confidence interval around the results obtained when using randomized galaxy redshifts. Absorbers with  log$_{10}(\tau_{\rm Ly\alpha})\gtrsim-1.0$ tend to be closer to galaxies than a random place in the Universe is, while lower optical depth pixels tend to be further away. }\label{NearestGalaxy}
\end{figure}

Figure~\ref{NearestGalaxy} shows the median 3-D Hubble distance to the nearest galaxy in our sample as a function of a pixel's Ly$\alpha$ optical depth, normalized by the median 3-D Hubble distance to the nearest galaxy for all pixels ($2.88$~pMpc), irrespective of their optical depth (the unnormalized distance is shown on the right $y$-axis). Note that because we divided by the median distance for all pixels, the result is not directly dependent on the completeness of our sample (the actual, unnormalized distances are, however, highly sensitive to the completeness). The error bars were determined by bootstrap resampling the QSO spectra using chunks of 500 pixels. The grey shaded area shows the $1\sigma$ confidence interval of results obtained for randomized galaxy samples. It shows what would be expected if absorbers and galaxies were distributed randomly with respect to each other. 

We cut the plot at the low optical depth end at the value corresponding to a $2\sigma$ detection of absorption for a S/N of 50. Noise prevents us from distinguishing lower optical depths from each other, causing the trend to flatten. At the high optical depth end we cut the plot at $\tau_{\rm Ly\alpha} = 10$. Although we can recover higher optical depths, we find that pixels with  $\tau_{\rm Ly\alpha} > 10$ are biased: while the median redshift stays close to 2.36 for optical depths in the plotted range, it increases to about 2.5 for higher optical depths. Because of our selection function, the galaxy number density is lower for such high redshifts, causing an upturn of the distance to the nearest galaxy for $\tau > 10$. 

The optical depth is strongly anti-correlated with the distance to the nearest galaxy. Pixels with $\tau> 0.1$, are closer to galaxies than  a random place in the Universe is, while pixels with $\tau <0.1$ are on average farther from galaxies than a random location is.

To provide some physical interpretation of this observed trend we use the Jeans approximation described in Section~\ref{Interpretation} to convert optical depths into estimates for the gas overdensities (implicitly smoothed on the scale of the absorbers, i.e.\ $\sim 10^2$~pkpc) (top $x$-axis). Gas at the mean density of the Universe would produce log$_{10}\tau_{\rm Ly\alpha}\approx -0.65$ and is thus typically closer to galaxies than a random place in the Universe. Although it is unclear whether we can estimate the overdensity sufficiently accurately for this statement to be reliable, it would in fact not be a surprising result. Because voids take up most of the volume, a random place in the universe will be underdense. 

If we cube the normalized distance to the nearest galaxy, then we get something close to the inverse of the overdensity of galaxies. However, in that case the scale over which this number density is measured, would be the distance to the nearest galaxy and would thus vary with the density itself. It is more instructive to measure the overdensity of galaxies on a fixed length scale, which is something we can also measure from our data. 

\begin{figure}
\includegraphics[height=3in,trim=1.5cm 0.5cm 3cm 1.5cm,clip]{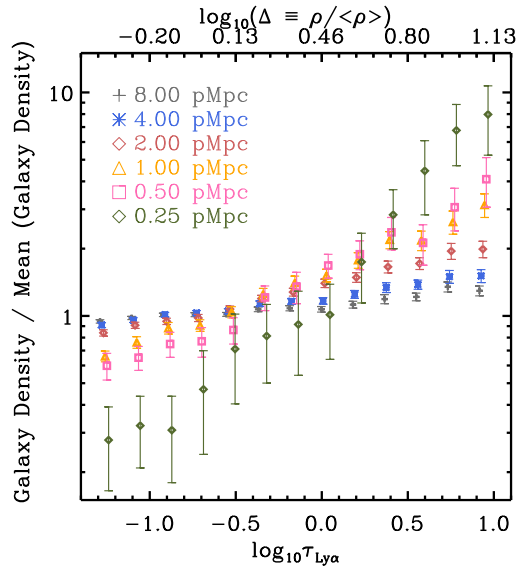}
\caption{The mean overdensity of galaxies in our sample within 0.25 (green diamonds), 0.5 (pink squares), 1 (orange triangles), 2 (red diamonds), 4 (blue asterisks), and 8 (black crosses) 3-D Hubble pMpc from pixels of a given \lya\ optical depth. The mean galaxy density is higher near stronger absorbers, and the density contrast is more pronounced when estimated on smaller scales. The data from this figure is tabulated in Table~\ref{tbl-4}.}\label{MeanGalaxyDensity}
\end{figure}

In Figure~\ref{MeanGalaxyDensity} we show the mean  overdensity of galaxies in our sample as a function of pixel optical depth, evaluated on (3-D Hubble) scales of 0.25, 0.5, 1, 2, 4, and 8 pMpc around them (tabulated in Table~\ref{tbl-4}). Note that these galaxy number densities cannot be directly compared with the gas overdensities that we can estimate from the optical depths because they are evaluated over different length scales. 

The galaxy number density increases with increasing optical depth, and the density contrast is stronger when it is measured on smaller scales. For example, pixels with $\tau_{\rm Ly\alpha}\sim 10$ (at $z\approx 2.4$) see on average a galaxy overdensity $\approx 2$ within 8 pMpc, but an overdensity $\sim 10$ within 0.25 pMpc. Pixels with $\tau_{{\rm Ly}\alpha} \sim 1$ typically reside in regions where the galaxy number density is close to the cosmic mean on scales $\ge 0.25$~pMpc. Because galaxies are clustered, this implies that for such pixels the distance to the nearest galaxy will typically be smaller than it is for a random place in the Universe (see Fig.~\ref{NearestGalaxy}). 

Given that we have both measurements of the galaxy number density and estimates of the gas overdensities, we could attempt to estimate the bias of these two components relative to each other by computing the ratio of the two densities. However, the length scale over which gas densities corresponding to Ly$\alpha$ absorbers are implicitly smoothed \citep[i.e.\ the local Jeans scale in case of overdense gas;][]{Schaye2001}, varies with the density and hence the optical depth, while galaxy and gas density have to be evaluated on the same scale in order to calculate the relative bias. We could in principle estimate the galaxy overdensity on scales that vary with the optical depth, but for overdense gas the relevant scales are somewhat too small to get a robust estimate of the galaxy number density. 

\begin{deluxetable*}{ccccccc}
\tabletypesize{\scriptsize}
\tablecaption{Data from Figure~\ref{MeanGalaxyDensity}. 
\label{tbl-4}}
\tablewidth{0pt}
\tablehead{\colhead{log$_{10}\tau_{\rm Ly\alpha}$\tablenotemark{a}} &	\colhead{$\Delta_{\rm g, 8 pMpc}$\tablenotemark{b} }& 	\colhead{$\Delta_{\rm g, 4 pMpc}$\tablenotemark{c}} & 	\colhead{$\Delta_{\rm g, 2 pMpc}$\tablenotemark{d}}& 	\colhead{$\Delta_{\rm g, 1 pMpc}$\tablenotemark{e}}& 	\colhead{$\Delta_{\rm g, 0.5 pMpc}$\tablenotemark{f}}	& 	\colhead{$\Delta_{\rm g, 0.25 pMpc}$\tablenotemark{g}}}
\startdata

-1.39 -- -1.21& 0.95$\pm{0.02}$&0.91$\pm{0.02}$&0.84$\pm{0.03}$&0.66$\pm{0.04}$&0.60$\pm{0.08}$&0.28$\pm{0.11}$\\
-1.21 -- -1.02 & 0.98$\pm{0.02}$&0.96$\pm{0.02}$&0.91$\pm{0.03}$&0.76$\pm{0.05}$&0.65$\pm{0.08}$&0.32$\pm{0.11}$\\
-1.02 -- -0.84 & 1.01$\pm{0.02}$&1.01$\pm{0.02}$&0.95$\pm{0.03}$&0.88$\pm{0.05}$&0.75$\pm{0.10}$&0.31$\pm{0.13}$\\
-0.84 -- -0.65 & 1.01$\pm{0.02}$&1.03$\pm{0.03}$&0.98$\pm{0.04}$&0.91$\pm{0.05}$&0.77$\pm{0.12}$&0.47$\pm{0.23}$\\
-0.65 -- -0.47 & 1.02$\pm{0.02}$&1.05$\pm{0.03}$&1.07$\pm{0.03}$&1.04$\pm{0.06}$&0.86$\pm{0.12}$&0.71$\pm{0.31}$\\
-0.47 -- -0.29 & 1.07$\pm{0.03}$&1.13$\pm{0.03}$&1.20$\pm{0.04}$&1.25$\pm{0.07}$&1.21$\pm{0.15}$&0.81$\pm{0.31}$\\
-0.29 -- -0.10 & 1.08$\pm{0.03}$&1.16$\pm{0.04}$&1.28$\pm{0.05}$&1.41$\pm{0.10}$&1.35$\pm{0.22}$&0.92$\pm{0.37}$\\
-0.10 -- 0.08 & 1.07$\pm{0.03}$&1.17$\pm{0.04}$&1.40$\pm{0.06}$&1.52$\pm{0.10}$&1.68$\pm{0.21}$&1.01$\pm{0.37}$\\
0.08 -- 0.26 & 1.12$\pm{0.04}$&1.25$\pm{0.05}$ &1.49$\pm{0.07}$&1.78$\pm{0.14}$&1.89$\pm{0.28}$&1.74$\pm{0.60}$\\
0.26 -- 0.45 & 1.19$\pm{0.06}$&1.34$\pm{0.07}$ &1.66$\pm{0.10}$&2.19$\pm{0.19}$&2.37$\pm{0.39}$&2.83$\pm{0.84}$\\
0.45 -- 0.63 & 1.22$\pm{0.05}$&1.38$\pm{0.07}$ &1.72$\pm{0.10}$&2.18$\pm{0.22}$&2.12$\pm{0.44}$&4.46$\pm{1.63}$\\
0.63 -- 0.82 & 1.35$\pm{0.07}$&1.50$\pm{0.10}$ &1.95$\pm{0.16}$&2.63$\pm{0.31}$&3.06$\pm{0.66}$&6.77$\pm{2.07}$\\
0.82 -- 1.00 & 1.30$\pm{0.07}$&1.51$\pm{0.10}$ &1.99$\pm{0.17}$&3.14$\pm{0.38}$&4.09$\pm{1.02}$&7.97$\pm{2.73}$\\
\enddata
\tablenotetext{a}{Bins in log$_{10}\tau_{\rm Ly\alpha}$.}
\tablenotetext{b}{The mean overdensity of galaxies in our sample within 8 3-D Hubble pMpc, with the 1 $\sigma$ confidence interval.}
\tablenotetext{c}{The mean overdensity of galaxies within 4 3-D Hubble pMpc.}
\tablenotetext{d}{The mean overdensity of galaxies within 2 3-D Hubble pMpc.}
\tablenotetext{e}{The mean overdensity of galaxies within 1 3-D Hubble pMpc.}
\tablenotetext{f}{The mean overdensity of galaxies within 0.5 3-D Hubble pMpc.}
\tablenotetext{g}{The mean overdensity of galaxies within 0.25 3-D Hubble pMpc.}
\end{deluxetable*}

\section{Summary \& conclusions}\label{Summary}
 In this paper we studied the galaxy-IGM interface at $z\approx2-2.8$ in the 15 fields of the Keck Baryonic Structure
Survey. Each KBSS field is centered on a very bright background QSO of which high S/N, high resolution spectra
have been obtained, and includes a densely-sampled galaxy survey optimized to sample the same
range of redshifts probed by the IGM measurements.  Our analysis has included a total
of 679 galaxies, the subset of the larger KBSS sample that lies in the appropriate range of redshifts
within 2pMpc of the corresponding QSO sightline. The redshifts of 71 of our galaxies were measured directly from nebular lines observed in the near-IR, while the remaining redshifts were derived from ISM absorption and/or \lya\ emission lines and the redshift calibration of \citet{Rakic2011}. The median redshift of our galaxies is 2.36.

We analyzed the spectra using the pixel optical depth method, which uses an automatic algorithm to recover the \lya\ optical depth in each pixel, making use of higher order Lyman lines to  recover the optical depth in pixels with saturated \lya\ absorption. Our analysis relied almost entirely on median statistics. Because the \lya\ optical depth is proportional to the column density of neutral hydrogen, its distribution is physically more relevant than that of the flux. 

Our results can be summarized as follows (as before, all distances are proper):
\begin{itemize}

\item We presented the first 2-D map of absorption around galaxies, showing the median \lya\ optical depth as a function of transverse and LOS separations. By converting velocity differences along the LOS into distances assuming pure Hubble flow, redshift space distortions show up as anisotropies in the absorption maps. 

\item Hydrogen Ly$\alpha$ absorption is strongly enhanced near galaxies. Within $\sim 10^2~$pkpc the median optical depth is $\gtrsim 2$ orders of magnitude higher than in a random place ($\tau_{\rm Ly\alpha} \gtrsim 10$ versus $< 10^{-1}$). 

\item Two types of anisotropies are clearly detected:
\begin{enumerate} 
\item On scales $\sim 30 -  200 ~\rm km\,s^{-1}$, or $0.1 - 1$ pMpc, the median absorption is significantly stronger, by $\sim 1-2$ orders of magnitude, along the LOS than in the transverse direction. We showed that although this ``finger of God" effect may be explained by galaxy redshift errors,  it is probably dominated by peculiar velocities of gas within and around the halos. 
\item On scales of $\sim 1.4 - 2.0$ pMpc the absorption is significantly stronger (with $> 3\sigma$ significance) in the transverse direction than along the LOS. We interpret this compression in the LOS direction as large-scale infall towards the potential wells occupied by the galaxies, i.e. the \citet{Kaiser1987} effect. To our knowledge, this is the first time that this signature of gravitational collapse has been detected for gas. 
\end{enumerate}

\item The median optical depth as a function of impact parameter (over $\pm165\,\rm km\, s^{-1}$ intervals around galaxies) drops by more than an order of magnitude going from $b < 0.13$ pMpc to $0.13<b<0.18$ pMpc and remains approximately constant thereafter out to at least 2~pMpc, where it is still a factor of two higher than in a random part of the spectrum. The sharp transition occurs at a radius that is similar to the virial radius of the halos that are thought to host the galaxies.

\item The median optical depth as a function of 3-D Hubble distance (i.e.\ the 3-D distance computed under the assumption of pure Hubble flow) shows no steep drop at $10^2$ pkpc. Instead, it decreases smoothly away from galaxies. The lack of a steep drop reflects the fact that 3-D Hubble distances of a few times $10^2$ pkpc mix strong absorption arising at impact parameters $< 10^2$ pkpc with the mostly weaker absorption at larger transverse separations.

\item Excess absorption is detected with $>3\sigma$ significance out to 2.0~pMpc in the transverse direction. In terms of the 3-D Hubble distance the significance is $>3\sigma$ out to 2.8 pMpc and $1.5\sigma $ for 2.8 to 4 pMpc.

\item The median optical depth near galaxies with a fixed impact parameter is highly variable, suggesting that the gas is clumpy. The scatter in the median optical depth at fixed impact parameter is $\sim0.75$ dex. Beyond 100 pkpc this is large compared with the excess median absorption.

\item Converting optical depths to column densities, we estimate that the median neutral hydrogen column density decreases from $\gtrsim 10^{14.5}~\rm cm^{-2}$ within $10^2$ pkpc to $< 10^{13} ~\rm cm^{-2}$ for larger impact parameters. We note that if we had selected  the strongest absorber within $10^2$ pkpc of each galaxy, instead of taking all the pixels within that distance into account, the median column density of so selected systems would have been substantially higher.

\item Converting optical depths into overdensities (implicitly smoothed on the scale of the absorbers, i.e.\ $\sim 10^2~$pkpc) by applying the ``Jeans" approximation \citep{Schaye2001}, the fluctuating Gunn-Peterson approximation \citep[e.g.][]{Rauch1997}, or a fit to a hydrodynamical simulations \citep{Aguirre2002} all yield similar results, with the Jeans approximation and the hydro simulation agreeing particularly well. We estimate that median gas overdensities decrease from $\gtrsim 10$ within $10^2$ pkpc to $\sim1$ at 1 pMpc. Beyond this impact parameter the overdensity asymptotes to values smaller than unity, which is expected because underdense regions dominate the volume and gas densities inferred from optical depths are volume rather than mass weighted.

\item The covering fraction of absorbers with median optical depth $\tau_{\rm Ly\alpha} > 1$ within $165~\rm km\,s^{-1}$ is $100 ^{+0}_{-32}\%$ for $b < 100$ pkpc and $66\pm 16\%$ for $b < 200$ pkpc. If we use the maximum rather than the median optical depth than the latter fraction increases to $89^{+11}_{-18}\%$.

\item By turning to a pixel-centric view, i.e.\ estimating the median distance to the nearest galaxy as a function of the pixel's optical depth, we show that absorbers with $\tau_{\rm Ly\alpha}>0.1$ are typically closer to galaxies 
than a random location in the Universe, while weaker absorbers are farther from galaxies compared to random.

\item The average number density of galaxies around absorbers increases with increasing absorption strength, and the galaxy density contrast is higher when it is evaluated on smaller scales. For example, on scales of 0.25 pMpc the galaxy overdensity increases from $\sim 10^{-0.5}$ at $\tau_{\rm Ly\alpha} = 10^{-1}$ to $\sim 1$ at  $\tau_{\rm Ly\alpha} = 1$ and to $\sim 10$ at $\tau_{\rm Ly\alpha} = 10$. When evaluated in spheres with radius of 2 pMpc the galaxy overdensity increases from $\approx 0.9$ at $\tau_{\rm Ly\alpha} \sim 10^{-1}$ to $\approx 2$ for $\tau_{\rm Ly\alpha} \sim 10$. Absorbers with $\tau_{\rm Ly\alpha}\sim 1$ reside in regions where the galaxy number density is close to the cosmic mean on scales $\ge 0.25$ pMpc. As most of the volume is underdense, such absorbers are nearer to galaxies than a random place in the Universe is.
\end{itemize}

This work could be improved considerably by reducing the redshift errors, which will be possible through
observations of nebular emission lines with the new multi-object
near-IR spectrograph Keck I/MOSFIRE \citep{McLean2010}. Such observations would also allow us to investigate whether the absorption varies with the properties of the galaxies such as their stellar masses or star formation rates. 

In a future paper we will present a comparison of our observational findings with cosmological, hydrodynamical simulations, which will aid the interpretation of the absorption profiles that we see around galaxies. In addition, \citet{Rudie2012}  present an analysis of the same data based on Voigt profile decompositions.

\acknowledgments
We are very grateful to Milan Bogosavljevi\'c, Alice
Shapley,  Dawn Erb,  Naveen Reddy, Max Pettini, Ryan Trainor,
 and David Law for their invaluable
contributions to the Keck Baryonic Structure Survey, without which
the results presented here would not have been possible. We also thank Ryan Cooke for his help with the continuum fitting of QSO spectra, and we thank the anonymous referee for a careful reading of the manuscript, and for valuable suggestions. This work was
supported by an NWO VIDI grant (OR, JS), by the US National Science
Foundation through grants AST-0606912 and AST-0908805,
and by the David and Lucile Packard Foundation (CCS). CCS acknowledges
additional support from the John D. and Catherine T.
MacArthur Foundation and the Peter and Patricia Gruber Foundation.
We thank the W.~M.~Keck Observatory staff for their assistance
with the observations.We also thank the Hawaiian people, as without
their hospitality the observations presented here would not have
been possible.

\appendix
\section{The effect of redshift errors}
\label{sec:zerrors}

As discussed in Section~\ref{Redshifts}, there are random errors in the redshift measurements of 
$\approx 130\, \rm km\, s^{-1}$ for those obtained from LRIS spectra, and $\approx 60\, \rm km\, s^{-1}$ for the NIRSPEC subsample. We expect that adding errors that are small compared to the redshift errors that are already present in the
data will have very little effect, whereas adding errors that are
large compared to the actual redshift errors should strongly degrade
the signal. 

\begin{figure*}
\resizebox{0.4\textwidth}{!}{\includegraphics{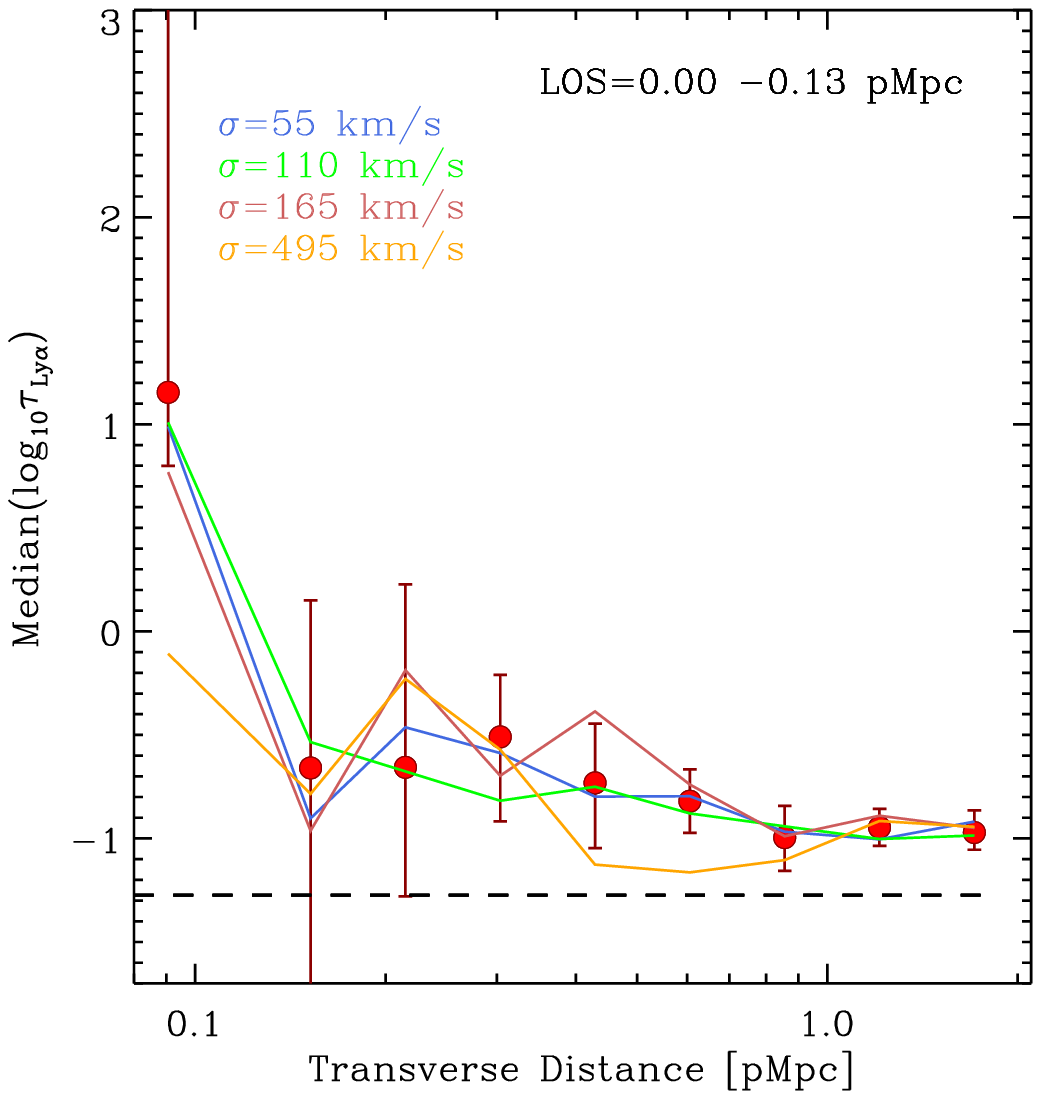}}
\resizebox{0.4\textwidth}{!}{\includegraphics{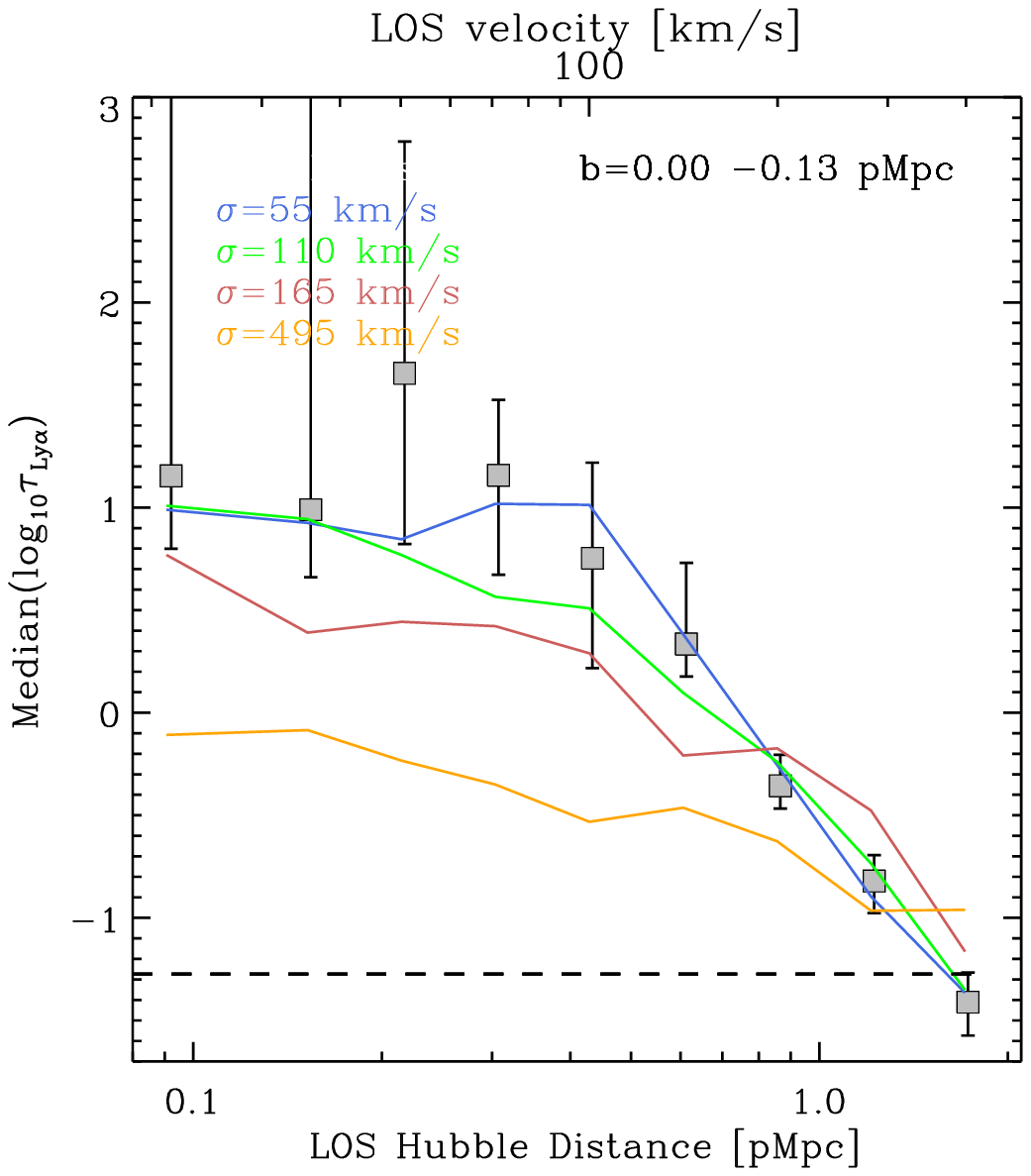}}
\caption{The effect of redshift errors. The data points with error
bars in left (right) panels shows the median log$_{10}\tau_{\rm
Ly\alpha}$ as a 
function of transverse (LOS) separation for the velocity (impact
parameter) range indicated in the panels. These correspond to the
first panel of Fig.~\ref{2Dcuts}. The blue, red, and orange curves show the result after adding random, Gaussian redshift errors with $\sigma=55$, 110, 165, and 495~$\rm km\, s^{-1}$, respectively. Significant differences start to appear for 165~$\rm km\, s^{-1}$, while three times greater redshift errors nearly completely wash out the signal. This suggests that the actual errors on the redshifts are $\lesssim 170\rm\, km\, s^{-1}$.}\label{LOSverror}
\end{figure*}

Figure~\ref{LOSverror} illustrates what effect the
addition of random, Gaussian redshift errors has. The left and right panels show median optical depth profiles in cuts along the $x$- and $y$-axes of Figure~\ref{2DdifferentialLOG} corresponding to the nearest LOS and transverse bins, respectively. The data points in the left (right) panel are thus identical to the red circles (black squares) of the first panel of Figure~\ref{2Dcuts}. 
The blue, green, red, and orange curves show the results after adding redshift errors with $\sigma = 55$, 110, 165, and 495 $\rm km\, s^{-1}$, respectively. This is done by adding a random error from a given Gaussian distribution to each galaxy redshift. Redshift errors smear out the signal along the LOS direction, thereby strongly reducing its amplitude. Adding 55  $\rm km\, s^{-1}$ errors hardly changes anything, but for 110 and 165 $\rm km\, s^{-1}$ significant differences start  to appear. Errors of 495 $\rm km\, s^{-1}$ completely wash out the signal. We conclude that the true redshift errors are $\lesssim 170\, \rm km\, s^{-1}$. 

\section{Correlations between data points}
\label{RankCorrelation}

Because a single galaxy contributes pixels to multiple distance bins along the LOS, we expect the errors to be correlated in the LOS direction. Conversely, since a single galaxy only has one impact parameter, we expect the errors to be uncorrelated in the direction transverse to the LOS. In this section we will show that this is indeed the case and that the errors along the LOS are correlated over scales of $\sim 10^2~\rm km \,s^{-1}$.

\begin{figure*}
\resizebox{0.4\textwidth}{!}{\includegraphics{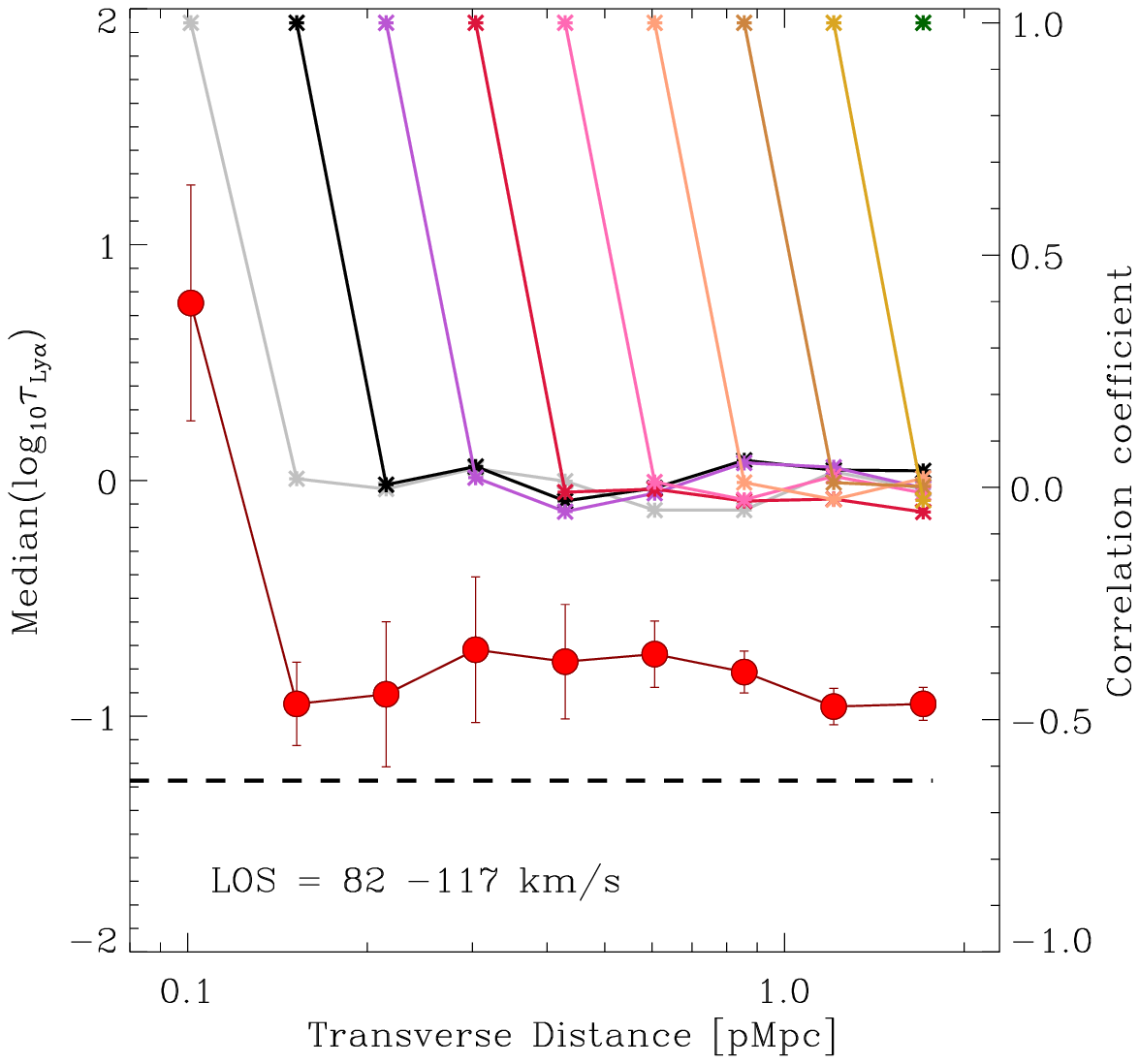}}\quad
\resizebox{0.4\textwidth}{!}{\includegraphics{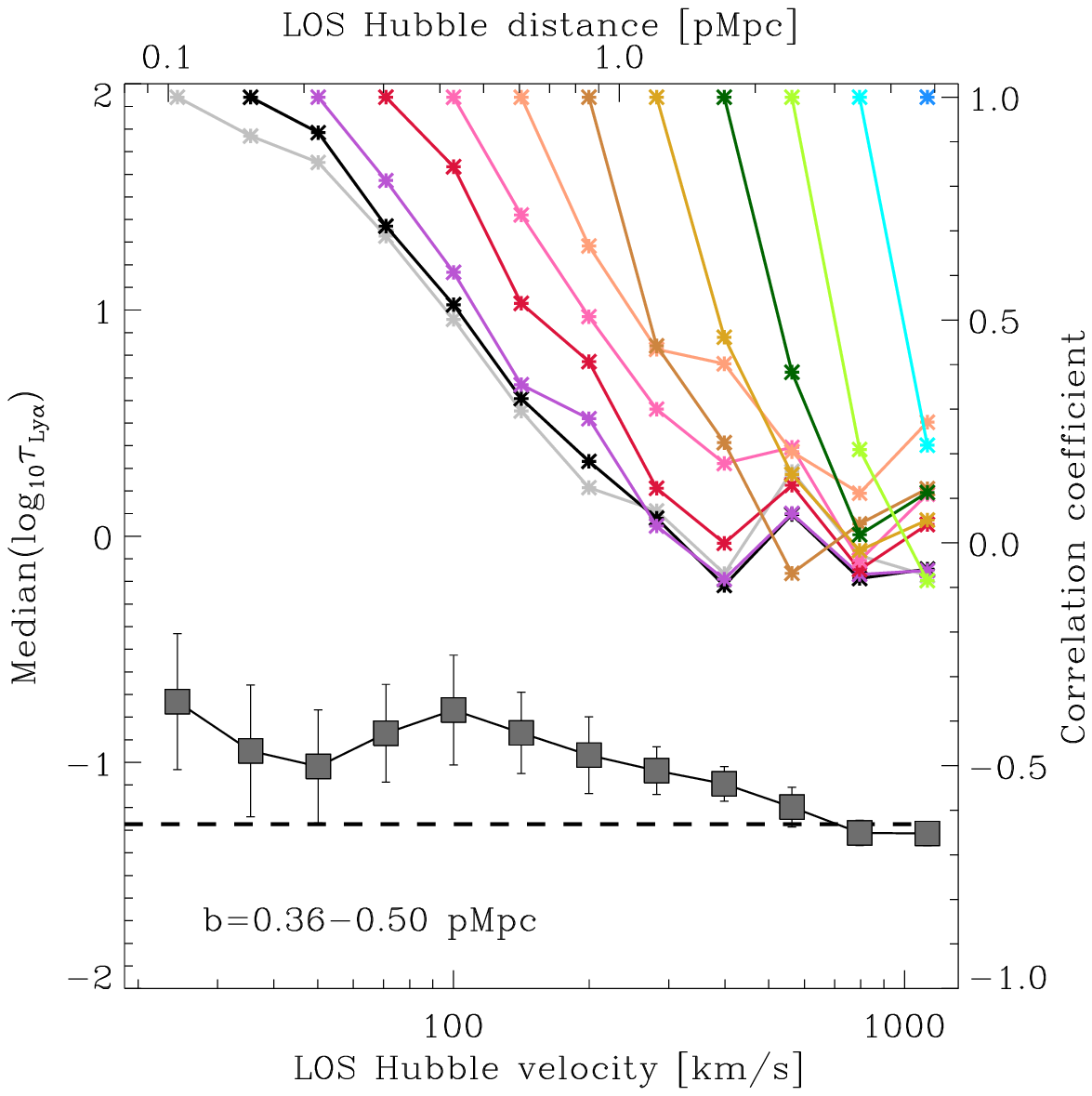}}
\caption{Correlations between errors of different data points. The data points with error bars in left (right) panels shows the median log$_{10}\tau_{\rm Ly\alpha}$ as a 
function of transverse (LOS) separation for the velocity (impact parameter) range indicated in the panels. These correspond to the central panel of Fig.~\ref{2Dcuts}. The different colored curves and the right $y$-axis show the Spearman rank correlation coefficient between the errors of different 
points, computed by correlating 1,000 different bootstrap samples (see text). Data points are uncorrelated in the transverse direction. Along the LOS they are strongly
correlated for separations $<100~\rm km\,s^{-1}$ but become
uncorrelated for distances $\gg 10^2~\rm km\,s^{-1}$.  }\label{ErrorCorrelation2D}
\end{figure*}

To study the correlation between points, we made 1,000 bootstrap realizations of our galaxy sample, found the median $\tau_{\rm Ly\alpha}$ in each of the distance bins for each of the realizations, and then computed the Spearman rank correlation coefficient between different distance bins. In other words, if we created a scatter plot of the median optical depth in bins $a$ and $b$ where each point is a bootstrap realization, then a high rank correlation would show up as points clustered along a line, whereas a correlation coefficient of zero would correspond to a random cloud of points in $\tau_a-\tau_b$ space.

We use rank correlation tests instead of the covariance matrix because some of the pixels are set to very high optical depths due to saturation, or to very low optical depths because they have an apparent positive flux (see Section~\ref{POD}). By using a non-parametric measure of the correlation strength, we are not susceptible to such problems (we use medians rather than averages for the same reason). 

The red circles in the left panel of Figure~\ref{ErrorCorrelation2D} shows the median optical depth for LOS separations $80-114~\rm km\,s^{-1}$ (i.e.\ 0.36 -- 0.50~pMpc for pure Hubble flow) as a function of transverse distance. Similarly, the black squares in the right panel show the median optical depth for transverse distances 0.36 -- 0.50~pMpc as a function of LOS separation. These points are thus identical to those shown in the central panel of Figure~\ref{2Dcuts}. 

The rest of the colored curves in Figure~\ref{ErrorCorrelation2D} show the rank correlation coefficient between different points, with the corresponding scale indicated on the right $y$-axis. Values of the correlation coefficient from 0.5 to 1 imply a strong correlation, between 0.2 and 0.5 a weak correlation, values around 0 imply there is no correlation, and negative values imply that there is an anti-correlation. 

For example, in the left panel the top red curve shows the correlation between the 4th point and all the ones that lie at larger distances. The rank correlation coefficient for the 4th point with itself is 1, but the correlation with the subsequent points is about zero. This confirms that the points are uncorrelated in the transverse direction.

On the other hand, the right panel shows that the points are strongly correlated for LOS separations $<100~\rm km\,s^{-1}$ and become uncorrelated for distances $\gg 10^2~\rm km\,s^{-1}$. This is consistent with the fact that we observe no evidence for structure along the LOS for separations $< 10^2~\rm km\,s^{-1}$ (see Figs.~\ref{2DdifferentialFixed} and \ref{2Dcuts}).

\begin{figure}
\resizebox{0.5\textwidth}{!}{\includegraphics{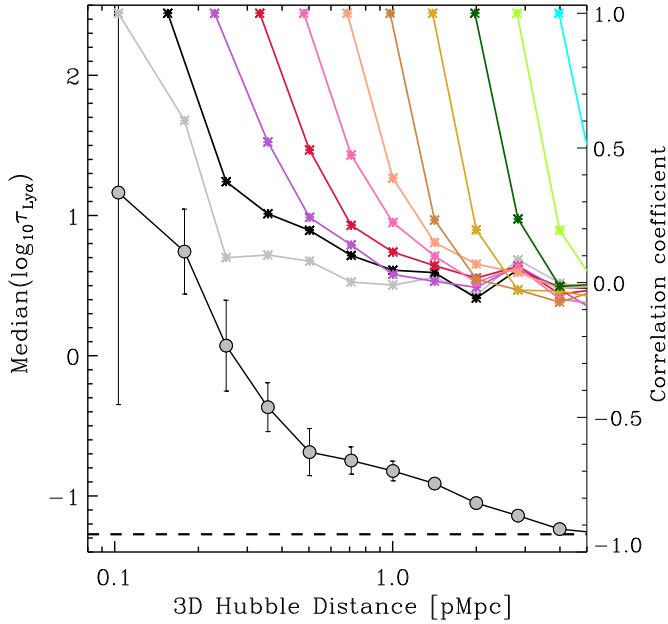}}
\caption{Similar to Figure~\ref{ErrorCorrelation2D} but showing the median optical depth and the rank correlation coefficient as a function of 3-D Hubble distance. The correlations are weaker than along the LOS (right panel of Fig.~\ref{ErrorCorrelation2D}) and are only substantial for adjacent bins.}
\label{ErrorCorrelation_LOG}
\end{figure}

Finally, Figure~\ref{ErrorCorrelation_LOG} shows the extent of the correlation between the errors of different points when we bin in terms of the 3-D Hubble distance. As the 3-D Hubble distance combines uncorrelated transverse with correlated LOS separations, the result must be intermediate between that for the two directions. Figure~\ref{ErrorCorrelation_LOG} demonstrates that this is indeed the case. Adjacent points are weakly correlated, while more distant points are essentially uncorrelated. 

\begin{thebibliography}{}
\bibitem[\protect\citeauthoryear{Adelberger et al.}{2003}]{Adelberger2003} Adelberger K. L.,  Steidel C. C., Shapley A. E., \& Pettini M., 2003, \apj, 584, 45
\bibitem[\protect\citeauthoryear{Adelberger et 
al.}{2004}]{Adelberger2004} Adelberger K.~L., Steidel C.~C., Shapley 
A.~E., Hunt M.~P., Erb D.~K., Reddy N.~A., \& Pettini M., 2004, \apj, 607, 226 
\bibitem[\protect\citeauthoryear{Adelberger et al.}{2005}]{Adelberger2005} Adelberger K. L., Shapley A. E., Steidel C. C., Pettini M., Erb D. K., \& Reddy N. A., 2005, \apj, 629, 636
\bibitem[\protect\citeauthoryear{Adelberger et al.}{2005b}]{Adelberger2005b} Adelberger K. L., Steidel C. C., Pettini M., Shapley A. E., Reddy N. A., \& Erb D. K. 2005, \apj, 619, 697
\bibitem[\protect\citeauthoryear{Aguirre et al.}{2002}]{Aguirre2002} Aguirre A., Schaye J., \& Theuns T., 2002, \apj, 576, 1
\bibitem[\protect\citeauthoryear{Barkana}{2004}]{Barkana2004} Barkana, R.\ 2004, \mnras, 347, 59
\bibitem[\protect\citeauthoryear{Becker et al.}{2011}]{Becker2011} Becker G.~D., Bolton J.~S., Haehnelt M.~G., \& Sargent W.~L.~W., 2011, MNRAS, 410, 1096 
\bibitem[\protect\citeauthoryear{Bergeron \& Boiss\'e}{1991}]{Bergeron1991} Bergeron J. \&  Boiss\'e P., 1991, \aap, 243, 344 
\bibitem[\protect\citeauthoryear{Bland \& Tully}{1988}]{Bland1988}{Bland} J.,  \& {Tully} B., 1988, Nature, 334, 43
\bibitem[\protect\citeauthoryear{Bolton et al.}{2005}]{Bolton2005} Bolton J.~S., Haehnelt M.~G., Viel M., \& Springel V., 2005, MNRAS, 357, 1178 
\bibitem[\protect\citeauthoryear{Bordoloi et 
al.}{2011}]{Bordoloi2011} Bordoloi R., et al., 2011,  ApJ, in press (arXiv:1106.0616)
\bibitem[\protect\citeauthoryear{Borthakur et al.}{2010}]{Borthakur2010}{Borthakur} S., {Tripp} T.~M., {Yun} M.~S., {Momjian} E., {Meiring} J.~D., {Bowen} D.~V., \& {York} D.~G., 2010, \apj, 713, 131
\bibitem[\protect\citeauthoryear{Bouche et al.}{2011}]{Bouche2011} 
Bouche N., et al., 2011, MNRAS, in press (arXiv:1107.4618) 
\bibitem[\protect\citeauthoryear{Bowen et al.}{2002}]{Bowen2002}{Bowen} D.~V., {Pettini} M.,  \& {Blades} J.~C., 2002, \apj, 580, 169
\bibitem[\protect\citeauthoryear{Brooks et al.}{2009}]{Brooks2009} 
Brooks A.~M., Governato F., Quinn T., Brook C.~B., \& Wadsley J., 2009, ApJ, 
694, 396 
\bibitem[\protect\citeauthoryear{Chabrier}{2003}]{Chabrier2003} Chabrier, G., 2003, \pasp, 115, 763
\bibitem[\protect\citeauthoryear{Chen et al.}{1998}]{Chen1998}{Chen} {H.-W.}, {Lanzetta} K.~M., {Webb} J.~K., \& {Barcons} X., 1998, \apj, 498, 77C
\bibitem[\protect\citeauthoryear{Chen et al.}{2001}]{Chen2001}{Chen} {H.-W.}, {Lanzetta} K.~M., \& {Webb} J.~K., 2001, \apj, 556, 158
\bibitem[Conroy et al.(2008)]{Conroy2008} Conroy, C., Shapley, 
A.~E., Tinker, J.~L., Santos, M.~R., \& Lemson, G.\ 2008, \apj, 679, 1192 
\bibitem[\protect\citeauthoryear{Cowie \& Songaila}{1998}]{CowieSongaila1998} Cowie L. L.,  \& Songaila A., 1998, \nat, 394, 44
\bibitem[Crain et al.(2010)]{Crain2010} Crain, R.~A., McCarthy, 
I.~G., Frenk, C.~S., Theuns, T., \& Schaye, J.\ 2010, \mnras, 407, 1403 
\bibitem[\protect\citeauthoryear{Crighton et 
al.}{2011}]{Crighton2011} Crighton N.~H.~M., et al., 2011, MNRAS, 414, 28 
\bibitem[\protect\citeauthoryear{Danforth \& Shull}{2008}]{DanforthShull2008} Danforth C.~W., \& Shull J.~M., 2008, \apj, 679, 194
\bibitem[\protect\citeauthoryear{Dekel et al.}{2009}]{Dekel2009} 
Dekel A., et al., 2009, Nature, 457, 451 
\bibitem[\protect\citeauthoryear{Ellison et al.}{2000}]{Ellison2000} Ellison S. L., Songaila A., Schaye J., \& Pettini M., 2000, \aj, 120, 1175
\bibitem[Erb et al.(2006)]{Erb2006a} {Erb}, D.~K., {Shapley}, A.~E., {Pettini}, M., {Steidel}, C.~C.,
	{Reddy}, N.~A., \& {Adelberger}, K.~L., \ 2006, \apj, 644, 813 
\bibitem[\protect\citeauthoryear{Erb et al.}{2006b}]{Erb2006b}  {Erb}, D.~K., {Steidel}, C.~C., {Shapley}, A.~E., {Pettini}, M., {Reddy}, N.~A., \& {Adelberger}, K.~L., 2006, ApJ, 646, 107 
\bibitem[\protect\citeauthoryear{Erb et al.}{2006c}]{Erb2006c} {Erb}, D.~K., {Steidel}, C.~C., {Shapley}, A.~E., {Pettini}, M., {Reddy}, N.~A., \& {Adelberger}, K.~L., 2006, \apj, 647, 128
\bibitem[\protect\citeauthoryear{Faucher-Gigu{\`e}re et  al.}{2008}]{FG2008} Faucher-Gigu{\`e}re C.-A., Lidz A., Hernquist L., \& Zaldarriaga M., 2008, ApJ, 682, L9 
\bibitem[\protect\citeauthoryear{Faucher-Gigu{\`e}re \& Kere\v s}{2011a}]{FG2010} Faucher-Gigu{\`e}re, C.-A.,  \& {Kere\v s} D., 2011, \mnras, 412, L118 
\bibitem[\protect\citeauthoryear{Faucher-Gigu{\`e}re et al.}{2011b}]{FG2011} Faucher-Gigu{\`e}re, C.-A., Kere{\v s}, D., 
\& Ma, C.-P.\ 2011, \mnras, in press (arXiv:1103.0001)
\bibitem[\protect\citeauthoryear{Frank et al.}{2003}]{Frank2003}{Frank}, S., {Appenzeller}, I., {Noll}, S., \& {Stahl}, O., 2003, \aap, 407, 473
\bibitem[Fumagalli et al.(2011)]{Fumagalli2011} Fumagalli, M., Prochaska, J.~X., Kasen, D., et al.\ 2011, \mnras, 418, 1796 
\bibitem[\protect\citeauthoryear{Kaiser}{1987}]{Kaiser1987} Kaiser N., 1987, MNRAS, 227, 1 
\bibitem[\protect\citeauthoryear{Kacprzak et al.}{2011}]{Kacprzak2011} Kacprzak G.~G., Churchill C.~W., Evans J.~L., Murphy M.~T., 
\& Steidel C.~C.\ 2011, MNRAS, in press (arXiv:1106.3068)
\bibitem[\protect\citeauthoryear{Kere{\v s} et 
al.}{2005}]{Keres2005} Kere{\v s} D., Katz N., Weinberg D.~H., \& Dav{\'e} R., 2005, MNRAS, 363, 2 
\bibitem[\protect\citeauthoryear{Kim \& Croft}{2008}]{KimCroft2008} Kim, Y.-R., \& Croft, R.~A.~C.\ 2008, \mnras, 387, 377 
\bibitem[\protect\citeauthoryear{Kimm et al.}{2011}]{Kimm2011} Kimm T., Slyz A., Devriendt J., \& Pichon C., 2011, MNRAS, 413, L51 
\bibitem[\protect\citeauthoryear{Kollmeier et al.}{2003}]{Kollmeier2003} Kollmeier, J.~A., Weinberg, D.~H., Dav{\'e}, R., \& Katz, N.\ 2003, \apj, 594, 75 
\bibitem[\protect\citeauthoryear{Komatsu et al.}{2009}]{Komatsu2009} {Komatsu} E. et al., 2009, \apjs, 180, 330
\bibitem[\protect\citeauthoryear{Lanzetta \& Bowen}{1990}]{Lanzetta1990}{Lanzetta} K.~M., \& {Bowen} D., 1998, \apj, 357, 321
\bibitem[\protect\citeauthoryear{Lanzetta et al.}{1995}]{Lanzetta1995} {Lanzetta} K.~M., {Bowen} D.~V., {Tytler} D., \& {Webb} J.~K., 1995, \apj, 442, 538
\bibitem[\protect\citeauthoryear{Lehnert et al.}{1999}]{Lehnert1999} {Lehnert} M.~D., {Heckman} T.~M., \& {Weaver} K.~A., 1999, \apj, 523, 575
\bibitem[\protect\citeauthoryear{Lidz et al.}{2010}]{Lidz2010} Lidz A., Faucher-Gigu{\`e}re C.-A., Dall'Aglio A., McQuinn M., Fechner C., Zaldarriaga M., Hernquist L., \& Dutta S., 2010, ApJ, 718, 199 
\bibitem[\protect\citeauthoryear{McLean et al.}{2010}]{McLean2010} McLean, I.~S., Steidel, 
C.~C., Epps, H., et al.\ 2010, \procspie, 7735, 47
\bibitem[\protect\citeauthoryear{Ocvirk et al.}{2008}]{Ocvirk2008} Ocvirk P., Pichon C., \& Teyssier R., 2008, MNRAS, 390, 1326 
\bibitem[Padmanabhan(2002)]{Padmanabhan2002} Padmanabhan, T.\ 2002, Theoretical Astrophysics - Volume 3, Galaxies and Cosmology, by T.~Padmanabhan, pp.~638.~Cambridge University Press, December 
2002.~ISBN-10: 0521562422.~ISBN-13: 9780521562423  
\bibitem[\protect\citeauthoryear{Penton et al.}{2002}]{Penton2002}{Penton} S.~V., {Stocke} J.~T., \& {Shull} J.~M., 2002, \apj, 565, 720
\bibitem[\protect\citeauthoryear{Pieri et al.}{2006}]{Pieri2006}{Pieri} M.~M., {Schaye} J., \& {Aguirre} A., 2006, \apj, 638, 45
\bibitem[Prada et al.(2011)]{Prada2011} Prada, F., Klypin, A.~A., 
Cuesta, A.~J., Betancort-Rijo, J.~E., \& Primack, J.\ 2011, arXiv:1104.5130 
\bibitem[\protect\citeauthoryear{Prochaska et 
al.}{2011}]{Prochaska2011} Prochaska J.~X., Weiner B., Chen H.~-W., Mulchaey J.~S., \& Cooksey K.~L., 2011, arXiv:1103.1891 
\bibitem[\protect\citeauthoryear{Rakic et al.}{2011}]{Rakic2011} {Rakic}, O., {Schaye}, J., {Steidel}, C.~C., \& {Rudie}, G.~C., 2011, \mnras, 414, 3265 
\bibitem[\protect\citeauthoryear{Rauch et al.}{1997}]{Rauch1997} Rauch M., et al., 1997, ApJ, 489, 7 
\bibitem[\protect\citeauthoryear{Reddy et al.}{2008}]{Reddy2008}{Reddy}, N.~A., {Steidel}, C.~C.,  {Pettini}, M.,  {Adelberger}, K.~L.,  {Shapley}, A.~E., {Erb}, D.~K.,  \& {Dickinson}, M., 1998, \apjs, 175, 48
\bibitem[\protect\citeauthoryear{Rubin et al.}{2010}]{Rubin2009} Rubin K. H. R., Prochaska J. X., Koo D. C., Phillips A. C., \& Weiner B. J., 2010, \apj, 712, 574  
\bibitem[Rudie et al.(2012)]{Rudie2012} Rudie, G.~C., Steidel, 
C.~C., Trainor, R.~F., et al.\ 2012, arXiv:1202.6055, \apj, in press 
\bibitem[\protect\citeauthoryear{Ryan-Weber}{2006}]{Ryan-Weber2006}  Ryan-Weber E.~V., 2006, \mnras, 367, 1251
\bibitem[\protect\citeauthoryear{Schaye et al.}{2000a}]{Schaye2000a} 
Schaye J., Theuns T., Rauch M., Efstathiou G., \& Sargent W.~L.~W., 2000a, MNRAS, 318, 817 
\bibitem[\protect\citeauthoryear{Schaye et al.}{2000b}]{Schaye2000b} Schaye J., Rauch M., Sargent W. L. W., \& Kim T.-S., 2000b, \apjl, 541, 1
\bibitem[\protect\citeauthoryear{Schaye}{2001}]{Schaye2001} Schaye J., 2001, \apj, 559, 507
\bibitem[\protect\citeauthoryear{Schaye et al.}{2003}]{Schaye2003} Schaye J., Aguirre A., Kim T.-S., Theuns T., Rauch M., \& Sargent W. L. W., 2003, \apj, 596, 768
\bibitem[Schaye et al.(2007)]{Schaye2007} Schaye, J., Carswell, R.~F., \& Kim, T.-S.\ 2007, \mnras, 379, 1169 
\bibitem[\protect\citeauthoryear{Shapley et al.}{2005}]{Shapley2005} Shapley A.~E., Steidel C.~C., Erb D.~K., Reddy N.~A., Adelberger K.~L., Pettini M., Barmby P., \& Huang, J., 2005, \apj, 626, 698
\bibitem[\protect\citeauthoryear{Shone et al.}{2010}]{Shone2010} Shone, A.~M., Morris, 
S.~L., Crighton, N., \& Wilman, R.~J.\ 2010, \mnras, 402, 2520 
\bibitem[\protect\citeauthoryear{Simcoe et al.}{2006}]{Simcoe2006}{Simcoe}, R.~A., {Sargent}, W.~L.~W., {Rauch}, M., \& {Becker}, G., 2006, \apj, 637, 648
\bibitem[\protect\citeauthoryear{Songaila}{1998}]{Songaila1998} Songaila A., 1998, \aj, 115, 2184
\bibitem[\protect\citeauthoryear{Steidel \& Sargent}{1992}]{Steidel1992}{Steidel} C.~C., \& {Sargent} W.~L.~W., 1992, \apjs, 80, 1
\bibitem[\protect\citeauthoryear{Steidel et al.}{1994}]{Steidel1994} {Steidel} C.~C., {Dickinson} M., \& {Persson} S.~E., 1994, \apj, 437, 75
\bibitem[\protect\citeauthoryear{Steidel et al.}{1997}]{Steidel1997} {Steidel} C.~C., {Dickinson} M., {Meyer} D.~M., {Adelberger} K.~L., \& {Sembach} K.~R., 1997, \apj, 480, 568
\bibitem[\protect\citeauthoryear{Steidel et al.}{2003}]{Steidel2003} Steidel C. C., Adelberger K. L., Shapley A. E., Pettini M., Dickinson M., \& Giavalisco M., 2003, \apj, 592, 728
\bibitem[\protect\citeauthoryear{Steidel et al.}{2004}]{Steidel2004} Steidel C. C., Shapley A. E., Pettini M., Adelberger K. L., Erb D. K., Reddy N. A., \& Hunt M. P., 2004, \apj, 604, 534
\bibitem[\protect\citeauthoryear{Steidel et al.}{2010}]{Steidel2010} {Steidel} C.~C., {Erb} D.~K., {Shapley} A.~E., {Pettini} M., {Reddy} N., {Bogosavljevi{\'c}} M., {Rudie} G.~C., \& {Rakic} O., 2010, \apj, 717, 289 
\bibitem[\protect\citeauthoryear{Steidel et al.}{2011}]{Steidel2011} {Steidel}, C.~C.,  {Bogosavljevi{\'c}}, M.,  {Shapley}, A.~E.,  {Kollmeier}, J.~A.,  {Reddy}, N.~A.,  {Erb}, D.~K.,  \& {Pettini}, M., 2011, \apj, 736, 160
\bibitem[\protect\citeauthoryear{Stewart et al.}{2011}]{Stewart2011} Stewart K.~R., Kaufmann T., Bullock J.~S., Barton E.~J., Maller A.~H., Diemand J., \& Wadsley J., 2011, ApJ, 735, L1 
\bibitem[\protect\citeauthoryear{Trainor \& Steidel}{2012}]{Trainor2012} Trainor R., \& Steidel C. C., 2012, submitted to \apj
\bibitem[\protect\citeauthoryear{van de Voort et al.}{2011a}]{Voort2011a} van de Voort F., Schaye J., Booth C.~M., 
Haas M.~R., \& Dalla Vecchia C., 2011a, MNRAS, 414, 2458 
\bibitem[\protect\citeauthoryear{van de Voort et al.}{2011b}]{Voort2011b} van de Voort F., Schaye J., Booth C.~M., \& Dalla Vecchia C., 2011b, MNRAS, 415, 2782 
\bibitem[\protect\citeauthoryear{Vogt et al.}{1994}]{Vogt1994} Vogt S. S., Allen S. L., Bigelow B. C., Bresee L., Brown B., Cantrall T., Conrad A., \& Couture M., 1994, \procspie,  2198, 362
\bibitem[\protect\citeauthoryear{Wilman et al.}{2007}]{Wilman2007} Wilman, R.~J., Morris, 
S.~L., Jannuzi, B.~T., Dav{\'e}, R.,  \& Shone, A.~M.\ 2007, \mnras, 375, 735 
\end{thebibliography}
\end{document}